\documentclass[twocolumn,superscriptaddress]{revtex4-2}

\usepackage{graphicx} 
\usepackage{bm} 
\usepackage{color}
\usepackage{lipsum}
\usepackage{mathtools}
\usepackage{amsmath}
\usepackage{amssymb}
\usepackage{chemfig}
\usepackage{booktabs}

\begin{document}

\newcommand{\BL}[1]{{\color{blue} #1}}
\setcounter{page}{1} 

\title{Trade-offs between number fluctuations and response in nonequilibrium chemical reaction networks}

\author{Hyun-Myung Chun}
\email{hmchun@kias.re.kr}
\affiliation{School of Physics, Korea Institute for Advanced Study, Seoul, 02455, Korea}

\author{Jordan M. Horowitz}
\email{jmhorow@umich.edu}
\affiliation{Department of Biophysics, University of Michigan, Ann Arbor, Michigan, 48109, USA}
\affiliation{Center for the Study of Complex Systems, University of Michigan, Ann Arbor, Michigan 48104, USA}
\affiliation{Department of Physics, University of Michigan, Ann Arbor, Michigan 48109, USA}

\date{\today}

\begin{abstract}
    We study the response of chemical reaction networks driven far from equilibrium to logarithmic perturbations of reaction rates.
    The response of the mean number of a chemical species is observed to be quantitively limited by number fluctuations as well as  the maximum thermodynamic driving force.
    We prove these trade-offs for linear chemical reaction networks and a class of nonlinear chemical reaction networks with a single chemical species.
    Numerical results for several model systems support the conclusion that these trade-offs continue to hold for a broad class of chemical reaction networks, though their precise form appears to depend sensitively on the deficiency of the network.
\end{abstract}

\maketitle 

\section{Introduction}
Driven networks of chemically reacting species display a rich phenomenology that allows them to operate as a scaffold for an array of functions, both within engineered setups as well as in living organisms.
In many of these situations, a fruitful method for assessing performance is to measure how sensitively the network responds to external stimuli, and in particular how the supply of energy shapes that response.
Energy transduction can amplify sensitivity in biological signaling cascades~\cite{goldbeter1981amplified,qian2003thermodynamic,ge2008sensitivity,Mallory2019,owen2022ultra} and in gene transcription networks~\cite{Estrada2016,tran2018precision}, improving their efficacies. 
It can induce negative differential response that enhances robustness in biochemical networks~\cite{falasco2019negative}.
Moreover, response can serve as an order parameter to interrogate a dynamical nonequilibrium phase transition~\cite{nguyen2018phase}.
While the nonequilibrium nature of the driving is integral to these scenarios, most of our understanding to date about the general principles that shape response is limited to near-equilibrium situations.

Near thermodynamic equilibrium, response is dictated by fluctuations as encoded in the fluctuation-dissipation theorem (FDT)~\cite{callen1951irreversibility,kubo1966fluctuation,forster1975hydrodynamic,marconi2008fluctuation}.
In the present context, it relates the static (or zero-frequency) response of the mean number of a chemical species $\langle n\rangle$ due to a change in its chemical potential $\mu$ with the number variance ${\rm Var}\{n\}$:
\begin{equation}\label{eq:eqFDT}
\frac{\partial \langle n\rangle}{\partial\mu} = \beta {\rm Var}\{n\},
\end{equation}
where $\beta$ is the inverse temperature.
Sensitive networks must also be noisy; therefore, by driving the dynamics we can create functional networks that are low-noise and responsive.

The fundamental nature of the FDT suggests that response and fluctuations remain linked in some form even for systems driven far from equilibrium.
To date, a number of such relationships have been identified that link response to some measure of fluctuations~\cite{agarwal1972fluctuation,Harada2005,speck2006restoring,baiesi2009fluctuations,prost2009generalized,chetrite2009eulerian,seifert2010fluctuation,baiesi2013update}.
While the predictions offer fundamental insights, putting them to use requires system-specific information.
For some proposals, identifying the correct measure of fluctuations requires either knowledge of the steady-state distribution or information about the dynamics~\cite{Harada2005,Wang2016,Lubensky2010,caprini2021fdt} (see, e.g., \cite{baiesi2013update} for a review).
Others require an appropriate choice of perturbation~\cite{graham1977covariant,asheichyk2019response,dal2019linear,chun2021nonequilibrium}. 
An alternative approach is offered by relaxing the desire for equalities and instead investigating potential trade-offs (or inequalities).
For example, it was recently shown that the maximum response to any perturbation is generally limited by an information-theoretic measure of dynamical fluctuations ~\cite{dechant2020fluctuation}.
The inspiration for the present work is the recently established  thermodynamic-response inequalities for discrete and continuous Markov processes~\cite{owen2020universal,gao2022thermodynamic}.
Together these works offer hope that simple and general inequalities can be identified that can serve as guideposts for rationalizing response in nonequilibrium dynamics.

In this paper, we study fluctuation-response relations for chemical reaction networks (CRNs) in nonequilibrium steady states driven by thermodynamic forces.
Inspired by \cite{owen2020universal}, we consider specific classes of perturbations that have proven to lead to concrete predictions in the context of small stochastic systems, and analyze potential inequalities between the easily-measured quantities that traditionally appear in the FDT  \eqref{eq:eqFDT}: number response and fluctuations.
In particular, we have studied how the CRN responds to changes in the rate constants $k_{+\rho}$ and $k_{-\rho}$ of a chemical reaction channel, hypothesizing that there are simple trade-offs between number response, fluctuations, and the maximum thermodynamic driving force ${\mathcal F}$:
\begin{equation}\label{eq:main1}
    \left| k_{\pm\rho} \frac{\partial \langle n \rangle}{\partial k_{\pm\rho}} \right|
    \leq \alpha_1 {\rm Var} \{ n \},
\end{equation}
\begin{equation}\label{eq:main2}
    \left| k_{+\rho} \frac{\partial \langle n \rangle}{\partial k_{+\rho}} + k_{-\rho} \frac{\partial \langle n \rangle}{\partial k_{-\rho}} \right|
    \leq \alpha_2 {\rm Var} \{ n \} g(\mathcal{F}),
\end{equation}
where $\alpha_{1,2}$ are system-dependent prefactors, $g(\mathcal{F})$ is an increasing function with $g(0) = 0$, and the shorthand notation $k_{\pm \rho}$ means `$k_{+\rho}$ or $k_{-\rho}$' throughout.
The values of $\alpha_{1,2}$ and the form of $g(\mathcal{F})$, in general, depend on the topological structure of the CRN, the chemical species of interest, and the perturbed reaction.
We prove the validity of \eqref{eq:main1} and \eqref{eq:main2} with $g(\mathcal{F}) = \tanh(\mathcal{F}/4)$ for linear CRNs and for CRNs with birth-death dynamics.
For CRNs with zero deficiency, we further provide numerical evidence that supports the accuracy of \eqref{eq:main1} and \eqref{eq:main2} with  $g(\mathcal{F}) = \tanh(\mathcal{F}/4)$.
Numerical studies of CRNs with nonzero deficiency also suggest trade-offs between number response and fluctuations, though we have not been able to identify a functional form for $g({\mathcal F})$.
Our analysis builds on a rich literature regarding the topological structure of CRNs~\cite{gunawardena2003chemical,feinberg2019foundations} and how that interfaces with nonequilibrium thermodynamics, both in the context of macroscopic deterministic dynamics~\cite{nicolis1977self,de2011non,polettini2014irreversible,ge2016mesoscopic,rao2016nonequilibrium,avanzini2020thermodynamics,avanzini2021nonequilibrium} and within microscopic stochastic descriptions~\cite{schmiedl2007stochastic,rao2018conservation}.
These theoretical developments have already led to the identification of universal thermodynamic properties, such as fluctuation theorems~\cite{gaspard2004fluctuation,andrieux2004fluctuation,andrieux2007fluctuation}, as well as information-geometric trade-offs~\cite{yoshimura2021information,yoshimura2021thermodynamic}.

The fluctuation-response relations explored here are intimately related to the inequalities predicted in \cite{owen2020universal} for the response of nonequilibrium Markov jump processes; however, the trade-offs \eqref{eq:main1} and \eqref{eq:main2} are not a trivial consequence.
The theory developed in \cite{owen2020universal} applies only to a system with a finite number of microscopic states, and when applied to rate constant perturbations in CRNs leads to bounds that scale with system size.
However, the structure of the microscopic state space of a CRN is restricted by symmetry due to repeated subunits, not taken into account in previous work.
Consequently, the inequalities \eqref{eq:main1} and \eqref{eq:main2} can provide finite bounds on the response even in CRNs with an infinite number of microscopic states, and in macroscopic systems.
Thus, this work complements the previous analyses, demonstrating that inequalities predicted in \cite{owen2020universal} can be modified to a larger class of scenarios.

This paper is organized as follows.
In Sec.~\ref{sec:CRNs}, we set the stage by introducing notation.
In Sec.~\ref{sec:illustration}, we begin with a class of CRNs with a single chemical species to illustrate how the trade-offs arise from the topological structure of the microscopic state space.
The trade-offs for linear CRNs are proved in Sec.~\ref{sec:deficiency_zero}.
We also derive weaker inequalities between fluctuation and response for deficiency zero nonlinear CRNs.
In Sec.~\ref{sec:deficiency_zero_numerical}, we demonstrate the compact forms \eqref{eq:main1} and \eqref{eq:main2} may also hold for deficiency zero nonlinear CRNs by providing numerical results for various models.
In Sec.~\ref{sec:deficiency_nonzero}, we numerically examine the validity of the trade-offs \eqref{eq:main1} and \eqref{eq:main2} for deficiency nonzero CRNs.
We close the paper in Sec.~\ref{sec:conclusion} with a summary of the main results and concluding remarks.

\section{Chemical Reaction Networks}\label{sec:CRNs}

\begin{figure*}[t]
\centering
\includegraphics[width=2\columnwidth]{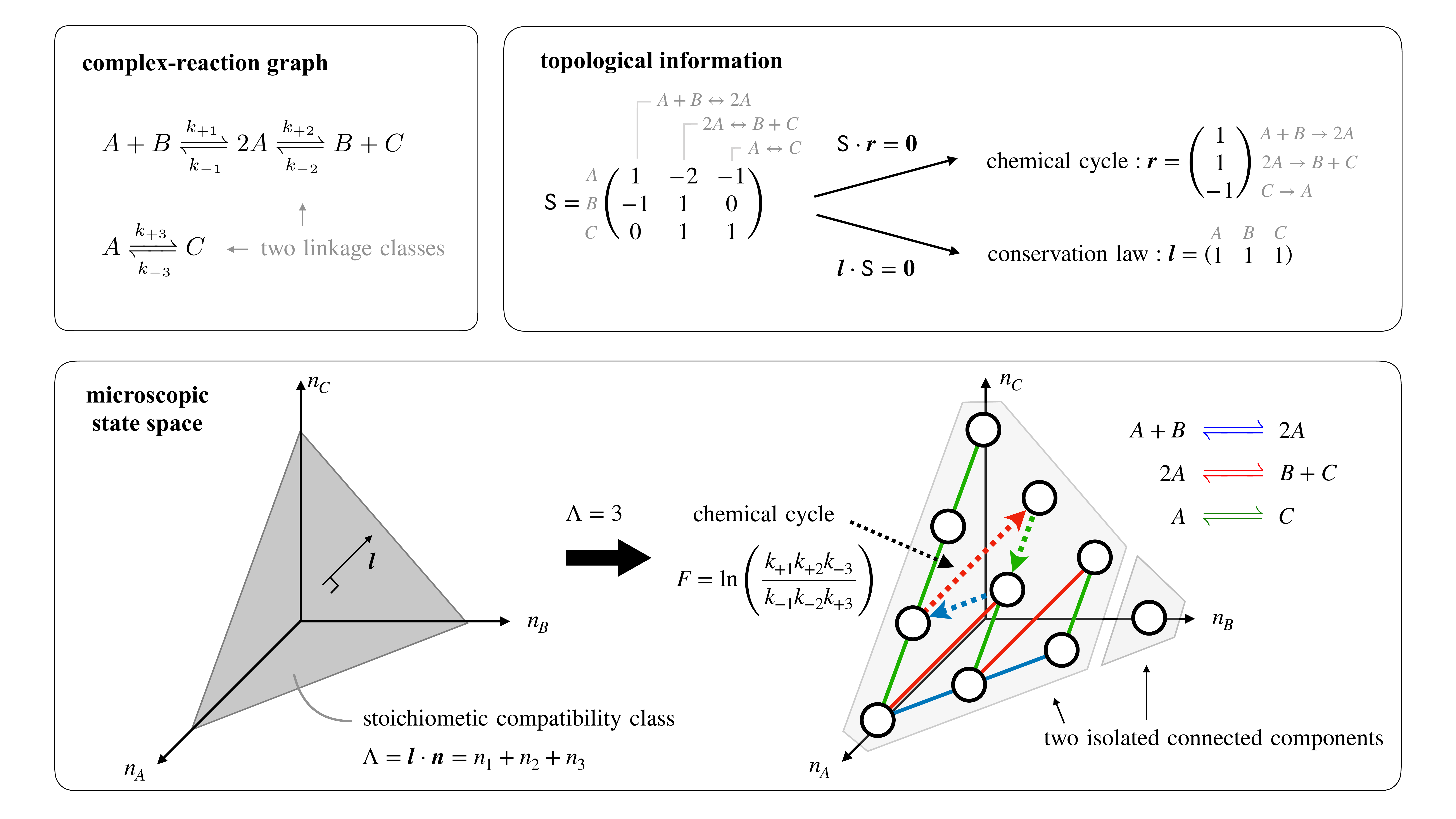}
\caption{Example of a chemical reaction network: It is characterized by three chemical species $\{X_1 = A, X_2 = B, X_3 = C\}$, five complexes $\{\bm{y}_1=(1,1,0)^{\rm T}, \bm{y}_2=(2,0,0)^{\rm T}, \bm{y}_3=(0,1,1)^{\rm T}, \bm{y}_4=(1,0,0)^{\rm T}, \bm{y}_5=(0,0,1)^{\rm T} \}$, and three reaction channels \{$\bm{\nu}_1^+ (= \bm{y}_1) \rightleftharpoons \bm{\nu}_1^- (= \bm{y}_2)$,  $\bm{\nu}_2^+ (= \bm{y}_2) \rightleftharpoons \bm{\nu}_2^- (= \bm{y}_3)$, $\bm{\nu}_3^+ (=\bm{y}_4)\rightleftharpoons \bm{\nu}_3^- (=\bm{y}_5)$\}.
The total number of chemical species $\bm{l} \cdot \bm{n} = n_1 + n_2 + n_3$ is conserved over time.}
\label{fig:example_CRN}
\end{figure*}

We begin by laying out the definitions and notations we will need to analyze response and noise in CRNs.  As a guide, we provide a concrete example of a CRN in Fig.~\ref{fig:example_CRN}.

\subsection{Constituents and Network Structure}

A CRN consists of $N_S$ chemical species, ${\bm X}=\{X_i\}_{i\in\mathcal{S}}$ with ${\mathcal S}=\{1,\dots, N_S\}$, reacting via $N_R$ reaction channels, $\rho \in \mathcal{R} = \{1, \cdots, N_R \}$,
\begin{equation}\label{eq:chem_reaction}
    {\bm \nu}_{\rho}^+ \cdot {\bm X}
    \rightleftharpoons
   {\bm \nu}_{\rho}^- \cdot {\bm X}.
   \end{equation}
Here, the elements of the stoichiometric vectors  ${\bm \nu}_{\rho}^+ = \{\nu_{i\rho}^+\}_{i\in\mathcal{S}}$  and ${\bm \nu}_{\rho}^- = \{\nu_{i\rho}^-\}_{i\in\mathcal{S}}$ are the stoichiometric coefficients that denote the number of species $X_i$ participating in the forward and reverse chemical reactions of each channel.
Note that the designation of forward or reverse (and reactant or product) is arbitrary, but fixed from the onset. 
Within this framework we can also include annihilation or creation due to exchanges with the surroundings by setting either ${\bm \nu}_{\rho}^+={\bm \emptyset}$ or ${\bm \nu}_{\rho}^-={\bm \emptyset}$, where ${\bm \emptyset} =(0,\dots,0)^{\rm T}$.
Lastly, to ensure thermodynamic consistency, we will assume all reaction channels are microscopically reversible, i.e., if a chemical reaction is present then so is its reverse.

Many of the features of the reaction dynamics will turn out to depend on the topological structure encoded in the CRN.  
We can exhibit this structure by collecting the changes caused by each reaction channel $\Delta\bm{\nu}_\rho={\bm\nu}_\rho^- - {\bm\nu}_\rho^+$ into the $N_S\times N_R$ stoichiometric matrix
\begin{equation}
    \mathsf{S} = \begin{pmatrix}
    \vert & \vert & & \vert \\
    \Delta\bm{\nu}_1 & \Delta\bm{\nu}_2 & \cdots & \Delta\bm{\nu}_{N_R} \\
    \vert & \vert & & \vert
    \end{pmatrix}.
\end{equation}
Each element $S_{i\rho} = \Delta\nu_{i\rho}$ represents the change in the number of species $X_i$ when reaction $\rho$ occurs. 
Denoting the kernel (or right null space) of $\mathsf{S}$ as $\ker(\mathsf{S})$, we then have by the rank-nullity theorem that only $s = N_R - \dim(\ker(\mathsf{S}))$ columns of $\mathsf{S}$ are linearly independent.
Without loss of generality, we can then reserve the first $s$ indices for linearly independent reactions: in other words, a vector $\Delta \bm{\nu}_\rho$ with $\rho > s$ is a linear combination of vectors $\Delta \bm{\nu}_\rho$ with $\rho \leq s$~\cite{polettini2014irreversible}.

The vectors in the left and right null space of $\mathsf{S}$ then have physical meanings that help to characterize the reaction dynamics~\cite{polettini2014irreversible,rao2016nonequilibrium,rao2018conservation}.
Vectors ${\bm l}$ in the left null space, ${\bm l}\cdot \mathsf{S}=0$, represent {\it conservation laws}: weighted combinations of chemical species that do not change when any reaction occurs.
Indeed, when reaction $\rho$ occurs the change in the quantity ${\bm l}\cdot{\bm X}$ is zero, ${\bm l}\cdot \Delta{\bm \nu}_\rho=0$.
We can then choose a set of linearly independent basis vectors $\{ \bm{l}_\alpha | \alpha = 1, \cdots, 
\dim(\ker(\mathsf{S}^{\rm T})) \}$, representing a set of independent conservation laws.
Similarly, a column vector $\bm{r}$ in the right null space, $\mathsf{S} \cdot \bm{r} = \bm{0}$, represents a {\it chemical cycle}: a series of reactions that leaves the numbers of all of chemical species unchanged.
Indeed, since the elements of $\mathsf{S}$ are all integers, the solution of the linear equation $\mathsf{S} \cdot \bm{r} = 0$ are rational numbers in general, which in turn can be rescaled to be integers.
In this way, a vector $\bm{r}$ can identify a series of reactions that do not change the numbers of each species, where each element ${\bm r}_\rho$ represents the number of occurrences of chemical reaction $\rho$ in the cycle and whose sign denotes the direction (see Fig.~\ref{fig:example_CRN}).
Again we can choose a set of linearly independent basis vectors $\{ \bm{r}_\alpha | \alpha = 1, 2, \cdots, \dim(\ker(\mathsf{S})) \}$.

Chemical reaction network theory has identified an alternative decomposition of the network structure that turns out to be important for characterizing the dynamics~\cite{gunawardena2003chemical,feinberg2019foundations}. 
Instead of analyzing the relationships between chemical species using the stoichiometric matrix, as we have done, we look at the relationships between the {\it complexes}: the collection of chemical species on either side of a reaction channel together with their stoichiometric coefficients.
For example, in the reaction channel $2X_1+X_2\rightleftharpoons 3X_3$, the complexes are $\{ 2X_1+X_2, 3X_3\}$ while the chemical species are $\{ X_1, X_2, X_3 \}$.
Each complex can then be identified with the corresponding stoichiometric vector ${\bm\nu}_\rho^\pm$.
For example, in the reaction $2X_1+X_2\rightleftharpoons 3X_3$, the complexes can be represented by the column vectors $(2,1,0)^{\rm T}$ and $(0,0,3)^{\rm T}$.
Now, not all $\bm{\nu}_\rho^\pm$ are distinct, so the number of complexes $N_C$ may be smaller than or equal to the number of chemical reactions ($2N_R$).
To emphasize this, we introduce a separate notation for the complexes $\bm{y}_l$  with $l \in \mathcal{C} = \{ 1, 2, \cdots, N_C\}$, collecting them into a $N_S\times N_C$ matrix ${\mathsf Y}$ whose elements contain information about which chemical species constitute each complex,
\begin{equation}\label{eq:matY}
    \mathsf{Y} = \begin{pmatrix}
    \vert & \vert &  & \vert \\
    \bm{y}_1 & \bm{y}_2 & \cdots & \bm{y}_{N_C} \\
    \vert & \vert &  & \vert
    \end{pmatrix}.
\end{equation}

The complexes further form a {\it complex-reaction graph}, where the complexes are the vertices and the reaction channel are the edges (see Fig.~\ref{fig:example_CRN}).
The topology of this graph is captured by the $N_C\times N_R$ incidence matrix $\mathsf{C}$, whose elements represent which reactions link which complexes, with signs denoting incoming or outgoing:
\begin{equation}
    C_{l\rho} = \begin{cases}
        1 & {\rm if} ~ \bm{y}_l = \bm{\nu}_\rho^-, \\
        -1 & {\rm if} ~ \bm{y}_l = \bm{\nu}_\rho^+, \\
        0 & {\rm otherwise}.
    \end{cases}
\end{equation}

Now, one of the central observations of the mathematical theory of CRNs is that the stoichiometric matrix can be decomposed as $\mathsf{S} = \mathsf{Y} \cdot \mathsf{C}$, linking how reactions change chemical species to how they change complexes.
An immediate consequence is the fact that the $\ker(\mathsf{S})$ may not coincide with $\ker(\mathsf{C})$, implying that there could be chemical cycles that are not visible in the complex-reaction graph.
This discrepancy can be captured by introducing the {\it deficiency} of the CRN $\delta = \dim(\ker(\mathsf{S}))-\dim(\ker(\mathsf{C}))$.  Its value has profound implications for what kinds of dynamical phenomenology the CRN can exhibit~\cite{feinberg2019foundations}.
When $\delta=0$, the dynamics has many features similar to equilibrium systems, whereas $\delta\neq 0$ has been linked to bistability~\cite{craciun2006bistable},  oscillations~\cite{tyson2008clocks}, and has been observed to affect the stochastic thermodynamics~\cite{polettini2015dissipation}.  We will also see that the CRN's deficiency has implications for nonequilibrium response as well.
The formula for deficiency we have introduced has a clear interpretation, but the original definition can be more useful for determining the value of $\delta$.
To this end, we note that a complex-reaction graph may be dissected into  $\ell$ isolated connected components called linkage classes.
Then, the deficiency of a CRN can be expressed as $\delta = N_C - \ell - s$~\cite{feinberg2019foundations}, which clearly shows it is a nonnegative integer.
For example, the CRN in Fig.~\ref{fig:example_CRN} has a deficiency of $\delta = 5 - 2 - 2 = 1$. 
It has no visible cycle in the complex-reaction graph but a single chemical cycle visible in the microscopic state space.

\subsection{Dynamics and Fluctuations}

To specify the dynamics, we consider a dilute, well-stirred mixture  in a fixed volume $\Omega$.
In this situation, an accurate dynamical model is provided by assuming mass-action kinetics, i.e., the rate of a chemical reaction is proportional to the concentration of the reactants. 
In other situations, distinct types of effective kinetics can emerge.
For example, Michaelis-Menten kinetics are widely used to model enzymatic reactions~\cite{cornish2013fundamentals}.
However, such emergent effective kinetics may be restored by taking appropriate limits from mass action kinetics~\cite{wachtel2018thermodynamically,avanzini2020thermodynamics}, and thus we do not explicitly consider them here.

Under our assumptions, an accurate description of the fluctuations is to model the chemical reactions as stochastic Poisson processes that randomly change the population of the chemical species $\bm{n} = (n_1, \cdots, n_{N_S})$, where $n_i$ denotes the number of species $X_i$.
Then by the law of mass action, the reaction rates are given by~\cite{gillespie1992rigorous,van2007stochastic}
\begin{equation}\label{eq:transition_rate}
    w_{\pm\rho}(\bm{n})
    = \frac{k_{\pm\rho}}{\Omega^{\sum_{i\in\mathcal{S}}\nu_{i\rho}^\pm - 1}} \frac{\bm{n}!}{(\bm{n} - \bm{\nu}_\rho^\pm)!},
\end{equation}
where the $k_{\pm\rho}$ are the rate constants, and $\bm{a}!$ is a shorthand for $\prod_{i\in\mathcal{S}} a_i!$.
As a result, the time-evolution of the probability $P(\bm{n},t)$ for the population to be $\bm{n}$ at time $t$ is given by the Chemical Master Equation~\cite{gillespie1992rigorous,van2007stochastic}
\begin{equation}\label{eq:CME1}
\begin{aligned}
    \partial_t P(\bm{n},t)
    & = \sum_{\rho\in\mathcal{R}}
    w_{+\rho}(\bm{n} - \Delta\bm{\nu}_\rho) P(\bm{n} - \Delta\bm{\nu}_\rho,t)  \\
    & ~~~ + \sum_{\rho\in\mathcal{R}}
    w_{-\rho}(\bm{n} + \Delta\bm{\nu}_\rho) P(\bm{n} + \Delta\bm{\nu}_\rho,t) \\
    & ~~~ - \sum_{\rho\in\mathcal{R}} \{ w_{+\rho}(\bm{n}) + w_{-\rho}(\bm{n}) \} P(\bm{n},t) \\
    & = \hat{\mathcal{L}} P(\bm{n},t).
\end{aligned}
\end{equation}
Note that conservation laws restrict the set of available states  where these dynamics can take place.
Indeed, recall that any basis vector ${\bm l}_\alpha$ in the left null space of ${\mathsf S}$ is by definition orthogonal to every $\Delta\bm{\nu}_\rho$.
Thus, chemical reactions, which change $\bm{n}$ only in increments of $\pm\Delta\bm{\nu}_\rho$, can only move the population through a subspace  orthogonal to the left null space of $\mathsf{S}$.
This subspace of accessible states is referred to as a (nonnegative) stoichiometric compatibility class and is identified by one particular set of values $\{ \Lambda_\alpha = \bm{l}_\alpha \cdot \bm{n} | \alpha = 1, 2, \cdots, \dim(\ker(\mathsf{S}^{\rm T})) \}$~\cite{feinberg2019foundations}.
Note also, that a stoichiometric compatibility class may consist of several isolated connected components.
We can visualize this microscopic state space as a graph(s) where the vertices are possible populations of the chemical species ${\bm n}$ and the edges represent allowable chemical reactions that connect accessible configurations (see Fig.~\ref{fig:example_CRN}).

We will assume that $P(\bm{n},t)$ converges to a steady-state distribution $\pi(\bm{n})$ in the long time limit,  given by the solution of $\hat{\mathcal{L}} \pi(\bm{n}) = 0$.
The total probability of the population being in an isolated connected component $\Gamma$, $p_\Gamma = \sum_{\bm{n}\in \Gamma} P(\bm{n},0)$, is conserved.
This constraint implies that the steady-state distribution segregates onto each component, $\pi(\bm{n}) = \sum_\Gamma p_\Gamma \pi_\Gamma(\bm{n})$, with conditional steady-state distribution $\pi_\Gamma(\bm{n})$ that we assume to be unique.
With knowledge of this distribution, we can characterize the steady-state fluctuations in the species numbers ${\bm n}$ using the mean and the variance
\begin{align}
\langle n_i \rangle &= \sum_{\bm{n}} n_i \pi(\bm{n})\\
{\rm Var}\{ n_i \} &= \sum_{\bm{n}} (n_i - \langle n_i \rangle)^2 \pi(\bm{n}).
\end{align}

Now, while the Chemical Master Equation offers an accurate description of the steady-state fluctuations, it can quickly become impractical to analyze, both numerically and analytically, as the system size grows, $\Omega\to\infty$.
In this macroscopic limit, the fluctuations are suppressed and  the limiting behavior of the populations is determined by the evolution of the concentrations of the chemical species, $\bm{n}/\Omega \to [\bm{X}]=( [X_1], \cdots, [X_{N_S}] )^{\rm T}$, where $[X_i]$ is the number of species $X_i$ per unit volume~\cite{kurtz1972relationship,gillespie2009deterministic}.
In this case, the dynamics are governed by the deterministic Chemical Rate Equation,
\begin{equation}\label{eq:CRE1}
\begin{split}
    \frac{d}{dt}[\bm{X}]
    & = \sum_{\rho \in \mathcal{R}}
    \Delta\bm{\nu}_\rho
    \left( k_{+\rho} [\bm{X}]^{\bm{\nu}^+_\rho} 
    - k_{-\rho} [\bm{X}]^{\bm{\nu}^-_\rho} \right)\\
    & = {\mathsf S}\cdot{\bm J}([{\bm X}])
    \end{split}
\end{equation}
where  $\bm{a}^{\bm{b}}$ is a shorthand for the component-wise product $\prod_{i\in\mathcal{S}} a_i^{b_i}$ with the convention $0^0 = 1$.
The column vector $\bm{J}([\bm{X}])$ has components $J_\rho([\bm{X}]) = k_{+\rho}[\bm{X}]^{\bm{\nu}_\rho^+} - k_{-\rho}[\bm{X}]^{\bm{\nu}_\rho^-}$ that 
represent the net reaction flux through each reaction channel.
We note that the same rate constants appear in the microscopic \eqref{eq:CME1} and the macroscopic dynamics \eqref{eq:CRE1} due to the inclusion of the volume-dependent scaling factor in the microscopic kinetics \eqref{eq:transition_rate}. 

Again, in the absence of intervention the concentrations will relax to a steady-state value $[\bm{X}]_{\rm ss}$ given as a solution of $\mathsf{S}\cdot \bm{J}([\bm{X}]_{\rm ss}) = \bm{0}$.
However, for a single isolated connected component of the microscopic state space, in sharp contrast to the unique steady-state distribution of the Chemical Master Equation~\eqref{eq:CME1}, the Chemical Rate Equation \eqref{eq:CRE1} may have multiple stable fixed points for certain collections of the rate constants, resulting in multiple potential steady-state values $[\bm{X}]_{\rm ss}$.
Which steady-state is observed as $t\to\infty$ depends on the initial conditions.
This distinction comes from the ordering of limits: $t\to \infty$ versus $\Omega \to \infty$.
Choosing the order of limits by taking $t \to \infty$ first, we can identify the unique steady-state concentration that is placed at the dominant peak of the probability distribution~\cite{vellela2009stochastic,ge2009thermodynamic}, which is one of the stable steady-state solutions of the deterministic equation \eqref{eq:CRE1} in multistable systems~\cite{ge2009thermodynamic}.
From now on this most-likely steady-state concentration will be simply denoted as $[\bm{X}]_{\rm ss}$, and we will only study the response and fluctuations around this configuration.

The Chemical Rate Equation is a deterministic equation, and therefore does not offer information about the fluctuations.
However, having identified the steady-state concentration $[{\bm X}]_{\rm ss}$, we measure the small fluctuations about this configuration from the scaled variance
\begin{align}
D_i=\lim_{\Omega\to\infty}\frac{1}{\Omega}{\rm Var}\{ n_i \},
\end{align}
which requires some knowledge of the microscopic dynamics.
In this article, we will use two methods to calculate these fluctuations in the macroscopic limit for different models.
When available, we will use the large deviation function $I([X]_i) = -\lim_{\Omega\to\infty}\Omega^{-1} \ln \pi(n_i = \Omega [X]_i)$ to analytically determine the scaled variance, via its curvature at $[{\bm X}]_{\rm ss}$ \cite{vellela2009stochastic}.
Otherwise, we will determine the fluctuations numerically using the Linear Noise Approximation to the Chemical Master Equation, which results in a computationally-tractable Langevin-type stochastic equation for the Gaussian fluctuations about $[{\bm X}]_{\rm ss}$~\cite{gillespie2000chemical,van2007stochastic}.
Details can be found in Appendix~\ref{sec:diffusion_approximation}.

\subsection{Steady-State Response}

For a fixed CRN, the steady-state mean number of chemical species $\langle {\bm n}\rangle$ is a function of the values of the rate constants.
Our concern here is how sensitively this average responds to changes to those rate constants.
As it will turn out  focusing on (dimensionless) logarithmic changes will facilitate the mathematical analysis.
In this case, the most generic perturbation is the linear combination
\begin{equation}
\frac{\partial\langle {\bm n}\rangle}{\partial \lambda} = \sum_{\rho \in {\mathcal R}} \lambda_{+\rho}k_{+\rho}\frac{\partial\langle {\bm n}\rangle}{\partial k_{+\rho}}+\lambda_{-\rho}k_{-\rho}\frac{\partial\langle {\bm n}\rangle}{\partial k_{-\rho}},
\end{equation}
with expansion coefficients $\lambda_{\pm\rho}$.
Similarly, in the macroscopic limit we can address the sensitivity of the steady-state concentration $\partial_\lambda [{\bm X}_{\rm ss}]$.

Making predictions for these arbitrary perturbations can be quite challenging.
Therefore inspired by the recent work  \cite{owen2020universal,gao2022thermodynamic} on the  response of nonequilibrium Markov processes, we will largely focus on two types of perturbations.
The first is perturbing a single rate constant logarithmically: $\partial/\partial\lambda = k_{\pm\rho}\partial/\partial k_{\pm\rho}$.
The second is a concerted change in the forward and reverse rate constants of a single reaction channel, which we will compactly denote as 
\begin{equation}\label{eq:symmetric_perturbation}
    \frac{\partial}{\partial B_{\rho}}
    = k_{+\rho} \frac{\partial}{\partial k_{+\rho}}
    + k_{-\rho} \frac{\partial}{\partial k_{-\rho}}.
\end{equation}
An example of such a perturbation is  a change in the concentration of a catalyst, which increases/decreases both rates of a reaction channel uniformly.

\subsection{Nonequilibrium Thermodynamics}

Driven or nonequilibrium chemical dynamics are characterized by the presence of nonzero reaction currents, a signal of time-reversal symmetry breaking.
This is only possible when there is an imbalance of reaction rates around chemical cycles, motivating the introduction of a thermodynamic force that measures the strength of this asymmetry.
Introducing a row vector $\bm{f}$ with components $f_\rho = \ln (k_{+\rho}/k_{-\rho})$, we define the {\it thermodynamic force} associated with the chemical cycle ${\bm r}$ by~\cite{rao2016nonequilibrium}
\begin{equation}
    F_{\bm r} = \bm{f} \cdot \bm{r}
    = \sum_{\rho\in\mathcal{R}} r_{\rho} \ln \left(\frac{k_{+\rho}}{k_{-\rho}}\right).
\end{equation}
When all the thermodynamic forces are zero, every reaction is balanced by its reverse, a statistical condition known as detailed balance true of all equilibrium systems.
Turning this around, we can say that a larger force corresponds to a system driven farther from equilibrium.
Note we identify the thermodynamic forces around chemical cycles (defined via the stoichiometric matrix), which should be distinguished from the loops in the microscopic state-space graph.
While these microscopic loops are used to identify thermodynamic forces in stochastic thermodynamics via Schnakenberg's network theory~\cite{schnakenberg1976network,rao2016nonequilibrium,rao2018conservation}, in the context of CRNs they are actually repeated copies of the chemical cycles, and therefore it is sufficient to consider forces just around the chemical cycles~\cite{polettini2014irreversible,rao2016nonequilibrium,rao2018conservation}.

To each reaction channel, we can associate a maximum thermodynamic force driving that current, 
\begin{equation}\label{eq:maxF}
{\mathcal F} = \max_{{\bm r}\ni \rho}|F_{\bm r}|
\end{equation} 
We will show that this quantity measures the strength of nonequilibrium driving through the reaction channel, and also constrains the response to changing the reaction channel's kinetic rates.

The physical meaning of thermodynamic forces becomes explicit once we take into account that our CRNs can be open to an environment containing other chemical species chemostatted to fix concentrations that facilitate the observed chemical reactions.
In this respect, the rate constants $k_{\pm \rho}$ introduced above should be considered pseudo-rate constants, having absorbed the concentrations of the chemostatted species into their definition.
While we assume that such a thermodynamically-consistent model can be constructed for the CRNs we study, we will not specify it in order to make the theory as general as possible.

\section{Birth-death process}\label{sec:illustration}

We begin our investigation with a class of CRNs with a single chemical species whose evolution is described by birth-death (or one-step) processes, defined by the property that every chemical reaction changes the number of chemical species by one~\cite{van2007stochastic}. 
This class includes CRNs that are both nonlinear and nonzero deficiency.
Despite this complication, there is a known closed-form expression for the steady-state probability distribution for general reaction rates, which will allow us to make concrete predictions for the limits of response in these models.

\begin{figure}[t]
\centering
\includegraphics[width=\columnwidth]{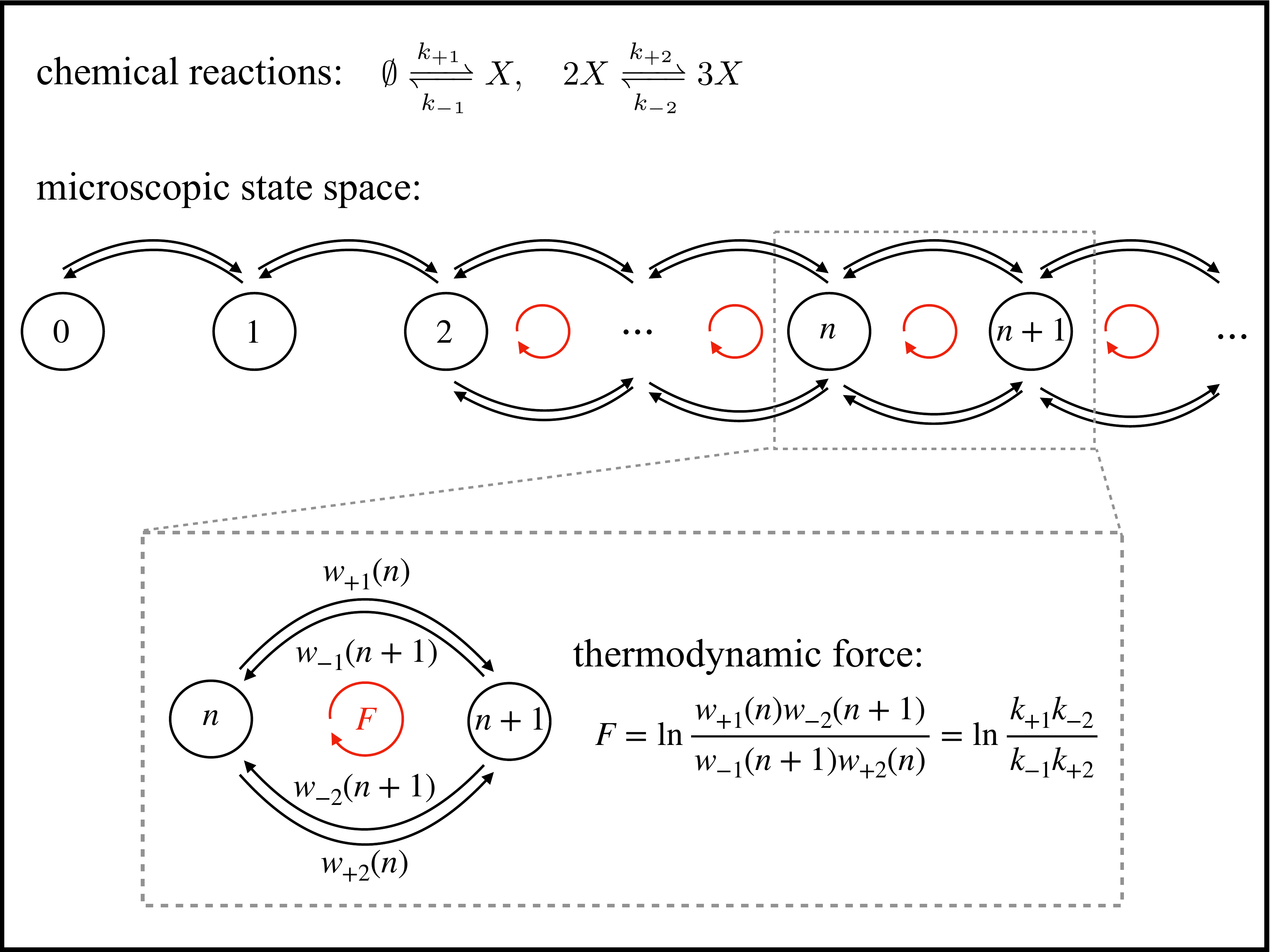}
\caption{Microscopic state space of the Schl\"ogl model and its thermodynamics.}
\label{fig:Schlogl_state_space}
\end{figure}

Consider the CRN composed of a single chemical species $X$ that reacts via the set of chemical reactions
\begin{equation}\label{eq:birth-death}
    z_\rho X \xrightleftharpoons[k_{-\rho}]{k_{+\rho}} (z_\rho+1) X
\end{equation}
with $\rho \in \mathcal{R} = \{ 1,2,\cdots,N_R \} $ and  $z_\rho$ are non-negative integers.
In turn, the reaction rates are given by
\begin{equation}
\begin{aligned}
    w_{+\rho}(n) & = \frac{k_{+\rho}}{\Omega^{z_\rho-1}} \frac{n!}{(n-z_{\rho})!}, \\
    w_{-\rho}(n) & = \frac{k_{-\rho}}{\Omega^{z_\rho}} \frac{n!}{(n-z_{\rho}-1)!}.
\end{aligned}
\end{equation}
The widely studied Schl\"ogl model fits into this framework with the choices $N_R = 2$, $z_1 = 0$, and $z_2 = 2$~\cite{schlogl1972chemical}.
We will use this model to illustrate the results of this section.  A depiction of its microscopic state space and thermodynamic force can be found in Fig.~\ref{fig:Schlogl_state_space}.

Owing to the simple network structure of the microscopic state space, exemplified in Fig.~\ref{fig:Schlogl_state_space}, we can find an exact solution to the steady-state distribution $\pi(n)$.
The key observation is that for a distribution to be stationary the reaction flux between any pair of neighboring states must be zero, $\sum_{\rho\in {\mathcal R}}w_{+\rho}(n)\pi(n) = \sum_{\rho\in {\mathcal R}}w_{-\rho}(n+1)\pi(n+1)$, which constrains the ratios of steady-state values all along the chain of states:
\begin{equation}\label{eq:pdf_ratio}
\frac{\pi(n+1)}{\pi(n)} = \frac{\sum_{\rho\in {\mathcal R}}w_{+\rho}(n)}{ \sum_{\rho\in {\mathcal R}}w_{-\rho}(n+1)}.
\end{equation}  
Iterating this observation using standard methodology (see Ref.~\cite{schlogl1972chemical})  leads to the steady-state distribution
\begin{equation}\label{eq:BD_steady-state_pdf}
    \pi(n) = \pi(0) \prod_{i=0}^{n-1} \frac{\sum_{\rho\in\mathcal{R}}w_{+\rho}(i)}{\sum_{\rho\in\mathcal{R}}w_{-\rho}(i+1)},
\end{equation}
where the probability of the boundary state $n=0$ is fixed by normalization $\sum_{n=0}^\infty \pi(n)=1$.

Although we have the steady-state distribution, finding a closed-form expression for the mean number $\langle n \rangle$ cannot be done in general without specifying the reaction rates.
So we analyze the steady-state response for an arbitrary perturbation directly, before specializing to the uni-directional and symmetric perturbations.
Using the identity $\sum_m \pi(m) \partial_\lambda \ln\pi(m) = 0$, which follows from probability conservation, we can rewrite the response $\partial_\lambda \langle n \rangle = \sum_n n \partial_\lambda\pi(n)$ as
\begin{equation}\label{eq:response_identity}
    \frac{\partial\langle n \rangle}{\partial \lambda}
    = \sum_{m,m'} m \pi(m)\pi(m') \frac{\partial}{\partial \lambda} \ln \frac{\pi(m)}{\pi(m')}
\end{equation}
shifting our attention to the sensitivity of ratios of probabilities, which will allow us to exploit the simple network structure of these models.
Indeed, taking advantage of \eqref{eq:pdf_ratio}, we have
\begin{equation}\label{eq:response_of_ratio}
\begin{aligned}
     \frac{\partial}{\partial \lambda} \ln \frac{\pi(m)}{\pi(m')}
    & = \sum_{i=m'}^{m-1} \frac{\partial}{\partial \lambda} \ln \frac{\pi(i+1)}{\pi(i)} \\
    & = \sum_{i=m'}^{m-1} \frac{\partial}{\partial \lambda} \ln\left( \frac{\sum_{\rho\in\mathcal{R}}w_{+\rho}(i)}{\sum_{\rho\in\mathcal{R}}w_{-\rho}(i+1)} \right)
\end{aligned}
\end{equation}
for $m>m'$.
The response for two arbitrarily distant states is given by the sum of the responses for neighboring states.

Let us first consider single-rate perturbations.
For two neighboring states, we have that
\begin{equation}\label{eq:BD_resp_ratio}
\begin{aligned}
    \left| k_{+\rho}\frac{\partial}{\partial k_{+\rho}} \ln \frac{\pi(n+1)}{\pi(n)} \right|
    & = \frac{w_{+\rho}(n)}{\sum_{\rho' \in \mathcal{R}}w_{+\rho'}(n)}\le 1 \\
    \left|k_{-\rho}\frac{\partial}{\partial k_{-\rho}} \ln \frac{\pi(n+1)}{\pi(n)} \right|
    & = \frac{w_{-\rho}(n+1)}{\sum_{\rho' \in \mathcal{R}}w_{-\rho'}(n+1)} \le 1,
\end{aligned}
\end{equation}
are bounded by 1.
Plugging \eqref{eq:BD_resp_ratio} into \eqref{eq:response_of_ratio} and summing over $i$, leads to the the inequality for non-neighboring states
\begin{equation}\label{eq:birth-death_ratio_bound}
    \left| k_{\pm\rho}\frac{\partial}{\partial k_{\pm\rho}} \ln \frac{\pi(m)}{\pi(m')} \right|
    \leq |m - m'|.
\end{equation}
The maximum response for the log-ratio of probabilities at two distant states is limited by the distance between them, i.e., the number of reactions required to move between the two configurations.
We then obtain a bound on \eqref{eq:response_identity} by first re-expressing the sum as 
\begin{equation}\label{eq:response_identity2}
\begin{split}
   & \left|k_{\pm\rho}\frac{\partial\langle n \rangle}{\partial k_{\pm\rho}}\right| \\
    &\ = \left|\sum_{m>m'} (m-m') \pi(m)\pi(m') k_{\pm\rho} \frac{\partial}{\partial k_{\pm\rho}} \ln \frac{\pi(m)}{\pi(m')}\right|\\
    &\ \le \sum_{m>m'} (m-m')^2 \pi(m)\pi(m'),
    \end{split}
\end{equation}
having bounded the absolute value using \eqref{eq:birth-death_ratio_bound} and that the sum is only over $m>m'$.
We recognize the result as 
\begin{equation}\label{eq:trade_off_BD1}
    \left| k_{\pm\rho} \frac{\partial \langle n \rangle}{\partial k_{\pm\rho}} \right|
    \leq {\rm Var}\{n\}.
\end{equation}
In words, large number fluctuations are necessary for high sensitivity.

Next, we consider the symmetric perturbation, which we anticipate will be constrained by the thermodynamic force in addition to the number fluctuations.
Again, motivated by \eqref{eq:response_of_ratio}, we start by analyzing the response of neighboring states.
Differentiating \eqref{eq:pdf_ratio}, we find
\begin{equation}\label{eq:sym_resp_BD1}
\begin{aligned}
    & \frac{\partial}{\partial B_{\rho}} \ln \frac{\pi(n+1)}{\pi(n)} \ \\
    & = \frac{w_{+\rho}(n)\bar{w}_{-\rho}(n+1) - w_{-\rho}(n+1)\bar{w}_{+\rho}(n)}{\{w_{+\rho}(n) + \bar{w}_{+\rho}(n)\} \{ w_{-\rho}(n+1) + \bar{w}_{-\rho}(n+1) \}}.
\end{aligned}
\end{equation}
where for convenience we have introduced the notation $\bar{w}_{\pm\rho}(n) = \sum_{\rho'\in\mathcal{R}\setminus\{\rho\}} w_{\pm\rho'}(n)$ for the total reaction rate not including the perturbed reaction channel.
Inspired by \cite{owen2020universal}, we factor the numerator as $(\sqrt{w_{+\rho}\bar{w}_{-\rho}} + \sqrt{w_{-\rho}\bar{w}_{+\rho}})(\sqrt{w_{+\rho}\bar{w}_{-\rho}} - \sqrt{w_{-\rho}\bar{w}_{+\rho}})$ and expand the denominator $(\sqrt{w_{+\rho}\bar{w}_{-\rho}} + \sqrt{w_{-\rho}\bar{w}_{+\rho}})^2 + (\sqrt{w_{+\rho}w_{-\rho}} - \sqrt{\bar{w}_{+\rho}\bar{w}_{-\rho}})^2$ to place the absolute-value of the response in a form that can readily be bounded:
\begin{equation}
    \left|  \frac{\partial}{\partial B_{\rho}} \ln \frac{\pi(n+1)}{\pi(n)} \right| = \frac{1}{1+a_n^2} \left|\tanh\left( \frac{\chi}{4}  \right)\right|
\end{equation}
with
\begin{equation}
    a_n = \frac{\sqrt{w_{+\rho}(n) w_{-\rho}(n+1) }
    - \sqrt{\bar{w}_{+\rho}(n) \bar{w}_{-\rho}(n+1) }}
    {\sqrt{w_{+\rho}(n) \bar{w}_{-\rho}(n+1) }
    + \sqrt{\bar{w}_{+\rho}(n) w_{-\rho}(n+1) }}.
\end{equation}
and 
\begin{equation}\label{eq:ratio_r}
   \chi =\ln\left( \frac{\sum_{\rho'\in\mathcal{R}\setminus\{\rho\}}w_{+\rho}(n)w_{-\rho'}(n+1)}{\sum_{\rho'\in\mathcal{R}\setminus\{\rho\}}w_{+\rho'}(n)w_{-\rho}(n+1)}\right).
\end{equation}
Using $1/(1+a_n^2)\le 1$ and the monotonicity of the hyperbolic tangent, we can now bound the response as
\begin{align}\label{eq:sym_resp_BD2}
  \left|  \frac{\partial}{\partial B_{\rho}} \ln \frac{\pi(n+1)}{\pi(n)} \right| \le \tanh\left( \frac{|\chi |}{4} \right).
\end{align}
The last step is to constrain $\chi$ by the thermodynamic forces driving the CRN out of equilibrium \eqref{eq:maxF}.
Importantly, the only forces that will matter are those around chemical cycles that contain the perturbed reaction channel (that is $r_\rho\neq 0$), which we denote as
\begin{equation}
    F_{\alpha} = \ln \left( \frac{w_{+\rho}(n)w_{-\alpha}(n+1)}{w_{+\alpha}(n)w_{-\rho}(n+1)} \right)
    = \ln \left( \frac{k_{+\rho}k_{-\alpha}}{k_{+\alpha}k_{-\rho}} \right).
\end{equation}
With this definition we can reinterpret \eqref{eq:ratio_r} as  
\begin{equation}
    |\chi| = \left|\ln \left( \frac{\sum_{\rho'\in\mathcal{R}\setminus\{\rho\}}w_{+\rho'}(n)w_{\rho}(n+1)e^{F_{\rho'}}}{\sum_{\rho'\in\mathcal{R}\setminus\{\rho\}}w_{+\rho'}(n)w_{-\rho}(n+1)} \right) \right|\le {\mathcal F},
\end{equation}
which is bounded by the maximum thermodynamic force $\mathcal{F} = \max_{\rho'}\{ |F_{\rho'}| \}$ through $\rho$.
From which we conclude that
\begin{align}\label{eq:sym_resp_BD3}
  \left|  \frac{\partial}{\partial B_{\rho}} \ln \frac{\pi(n+1)}{\pi(n)}  \right| \le \tanh\left( \frac{\mathcal F}{4} \right).
\end{align}

Following the same procedure carried out for single-rate perturbations,
we can now obtain a trade-off between fluctuations, response, and nonequilibrium driving,
\begin{equation}\label{eq:trade_off_BD2}
    \left| \frac{\partial \langle n \rangle}{\partial B_\rho} \right|
    \leq {\rm Var}\{n\} \tanh\left(\frac{\mathcal{F}}{4} \right).
\end{equation}
Note that this bound is tighter than a simple application of the single-rate bound, $|\partial\langle n \rangle/\partial B_\rho| \leq |k_{+\rho}\partial\langle n \rangle/\partial k_{+\rho} | + |k_{-\rho}\partial\langle n \rangle/\partial k_{-\rho} | \leq 2 {\rm Var}\{ n \}$.

Our analytic predictions also have implications for the limits to response in the macroscopic limit.
Dividing \eqref{eq:trade_off_BD1} and \eqref{eq:trade_off_BD2} by $\Omega$ and taking the $\Omega\to \infty$ limit, we can transfer our trade-offs to 
\begin{equation}\label{eq:trade_off_BD3}
    \left| k_{\pm\rho} \frac{\partial [X]_{\rm ss}}{\partial k_{\pm\rho}} \right| \leq D_X,
\end{equation}
\begin{equation}\label{eq:trade_off_BD4}
    \left| \frac{\partial [X]_{\rm ss}}{\partial B_\rho} \right| \leq D_X \tanh\left( \frac{\mathcal{F}}{4} \right),
\end{equation}
where $D_X = \lim_{\Omega\to \infty}{\rm Var}\{n\}/\Omega$ is the scaled variance.
Remarkably, the response of the \emph{macroscopic} steady-state concentration is limited by the \emph{microscopic} fluctuations, which are completely absent from the deterministic description.
While the information about the microscopic fluctuations becomes hidden in the macroscopic limit, its influence remains.
A similar observation was made recently in \cite{freitas2022emergent}, where the entropy production measured from macroscopic dynamics bounds the probability of rare fluctuations.

\begin{figure}[t]
\centering
\includegraphics[width=\columnwidth]{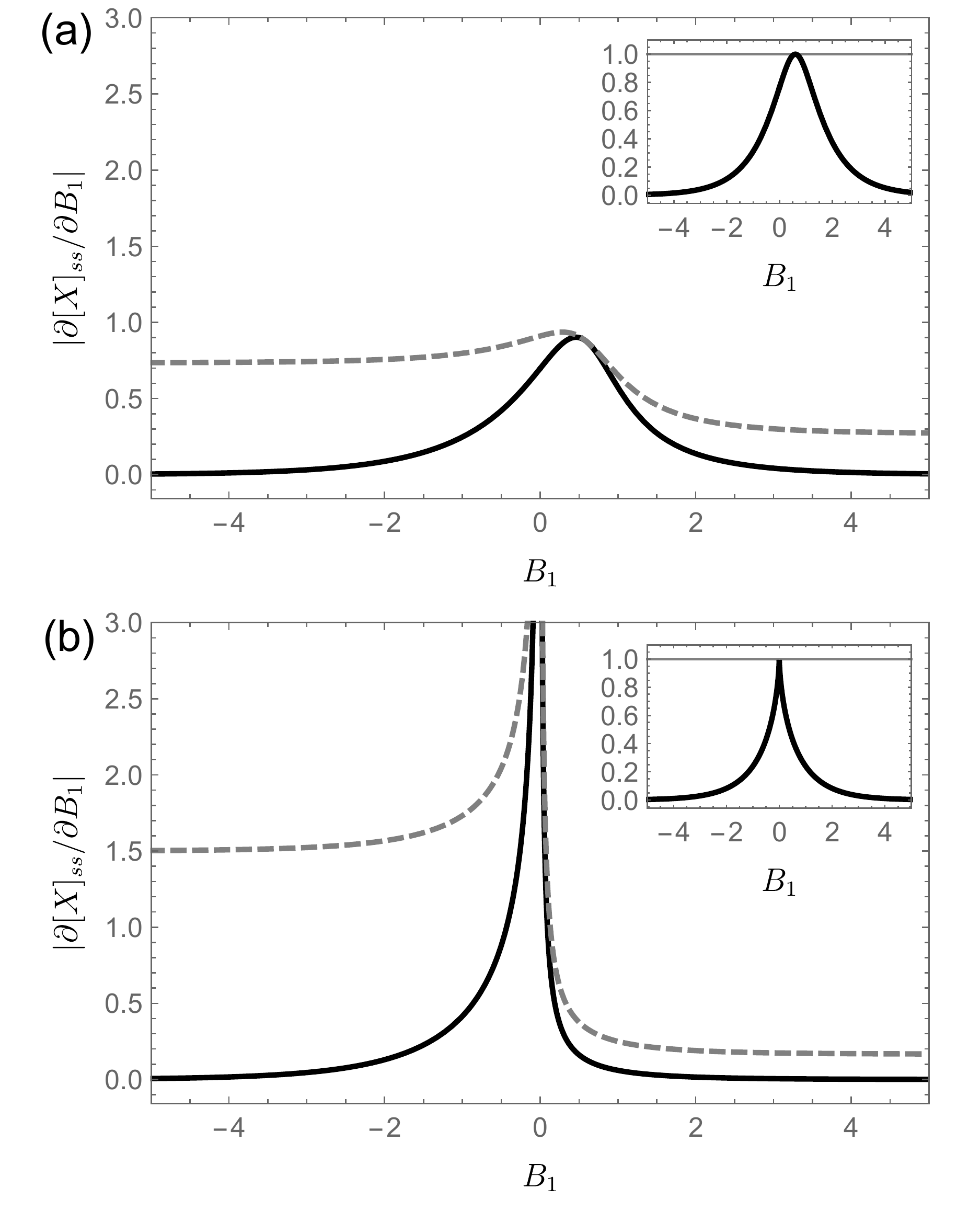}
\caption{The response of the steady-state concentration to a symmetric perturbation in the Schl\"ogl model.
The rate constants are  $k_{+1} = 9e^{B_1-F}$, $k_{-1}=3e^{B_1}$, $k_{+2} = 3$, and $k_{-2}=1$ with (a) $F = 1$ and (b) $F = \ln9$.
The dashed lines represent $D_X$ in (a) and $D_X \tanh(F/4)$ in (b).
The insets depict the ratio of the response to the bound.
The scaled variance $D_X$ is obtained from the second derivative of the large deviation function $I(x)=-\lim_{\Omega\to\infty}\Omega^{-1} \ln \pi(n=\Omega x)$~\cite{vellela2009stochastic}.
}
\label{fig:Schlogl_response}
\end{figure}

We validate the trade-off \eqref{eq:trade_off_BD4}, by numerically calculating the response and scaled variance of the Schl\"ogl model.
The steady-state concentration of the Schl\"ogl model satisfies the cubic equation
\begin{equation}\label{eq:cubic_eq_Schlogl}
    k_{+1} - k_{-1}x + k_{+2}x^2 - k_{-2}x^3= 0.
\end{equation}
Depending on the rate constants, the equation has one or three positive solutions.
We will focus on a regime of parameters where there is only a single fixed point to simplify the analysis.
Parameterizing the rate constants as $k_{+1} = 9e^{B_1-F}$, $k_{-1}=3e^{B_1}$, $k_{+2} = 3$, and $k_{-2}=1$, we find that the equation has a unique solution as long as $F\leq \ln 9$.
The response and number fluctuations are compared  in Fig.~\ref{fig:Schlogl_response} for $F<\ln 9$ and $F=\ln9$.
The bound disappears at the critical point where the number fluctuation diverges~\cite{nicolis1977stochastic}.
The ratio of the response to the bound demonstrates that inequality \eqref{eq:trade_off_BD4} is saturable (insets of Fig.~\ref{fig:Schlogl_response}).

\section{Deficiency Zero}\label{sec:deficiency_zero}

Trade-offs between response and fluctuation can also be shown to exist for deficiency zero CRNs.
We will distinguish two types of CRNs, linear and nonlinear, based on whether the dynamics in the deterministic Chemical Rate Equation is linear or nonlinear.  
For linear CRNs we will be able to derive simple bounds on the response in terms of number fluctuations, in line with \eqref{eq:main1} and \eqref{eq:main2}.
For nonlinear CRNs, we will provide numerical studies that suggest that number fluctuations constrain nonequilibrium response in these cases as well; however, 
we cannot prove them.
Some analytic progress is possible though, and other trade-offs between response and fluctuations will be demonstrated.

Deficiency zero CRNs have a  special property that makes their response to perturbations amenable to theoretical analysis.  
According to the deficiency zero theorem~\cite{feinberg2019foundations}, for every choice of rate constants, the fixed points of the Chemical Rate Equation \eqref{eq:CRE1} of a (weakly) reversible, deficiency zero CRNs are such that the net reaction flux between every pair of complexes perfectly balances, ${\mathsf C}\cdot {\bm J}([{\bm X}]_{\rm ss})={\bm 0}$, a property known as complex-balanced (cf.\ the general condition ${\mathsf S}\cdot{\bm J}([{\bm X}]_{\rm ss}) = {\mathsf Y}\cdot{\mathsf C}\cdot {\bm J}([{\bm X}]_{\rm ss}) ={\bm 0}$).
Furthermore, the complex-balanced steady-state concentration is unique up to a given stoichiometric compatibility class and stable.
There are a number of implications of this property that will turn out to be important for the study of response.

The first  is that the steady-state distribution of any complex-balanced CRN has a product form~\cite{anderson2010product,lubensky2010equilibriumlike}.
To be precise, for a state $\bm{n}$ on an isolated connected component $\Gamma$ of a stoichiometric compatibility class identified by a set of conservation laws $\{ \Lambda_\alpha = \bm{l}_\alpha^{\rm T} \cdot \bm{n} | \alpha = 1, 2, \cdots, \dim(\ker(\mathsf{S}^{\rm T})) \}$, the conditional steady-state probability is given by
\begin{equation}\label{eq:product_form}
    \pi_\Gamma(\bm{n})
    = \frac{1}{Z_\Gamma([\bm{X}]_{\rm ss})} \frac{(\Omega[\bm{X}]_{\rm ss})^{\bm{n}}}{\bm{n}!},
\end{equation}
where $Z_\Gamma([\bm{X}]_{\rm ss}) = \sum_{\bm{n} \in \Gamma} (\Omega[\bm{X}]_{\rm ss})^{\bm{n}}/\bm{n}!$ is a normalization factor.
Now generically if we were to change the rates of a complex-balanced CRN there is no guarantee that the resulting dynamics would be complex-balanced nor would  the steady-state distribution maintain its product form.
However, for deficiency zero CRNs, all steady-states are guaranteed to be complex-balanced, and so a perturbation only has the effect of shifting the complex-balanced steady-state concentrations while keeping the product form of  \eqref{eq:product_form} intact.
As a consequence, the static response of the conditional average number of any chemical species $\langle n_i \rangle_\Gamma = \sum_{\bm{n}\in\Gamma} n_i \pi_\Gamma(\bm{n})$ can be determined by differentiating \eqref{eq:product_form},
\begin{equation}\label{eq:FRR}
    \frac{\partial \langle n_i \rangle_\Gamma}{\partial \lambda}
    = \sum_{j\in\mathcal{S}} {\rm Cov}_\Gamma \{n_i, n_j \}  \frac{\partial \ln [X_j]_{\rm ss}}{\partial \lambda}, 
\end{equation}
where ${\rm Cov}_\Gamma \{n_i,n_j\} = \langle n_i n_j \rangle_\Gamma - \langle n_i \rangle_\Gamma \langle n_j \rangle_\Gamma$.
This equation \eqref{eq:FRR} links the microscopic stochastic response ($\partial_\lambda \langle n_i \rangle_\Gamma$) with the macroscopic deterministic response ($\partial_\lambda \ln[X_j]_{\rm ss}$) through the fluctuations (${\rm Cov}_\Gamma\{n_i,n_j\}$).

To make further progress, we need to constrain the macroscopic response, and again restricting attention to deficiency zero CRNs offers a method.
It turns out that the equation for the complex-balanced steady-state concentration ${\mathsf C}\cdot {\bm J}([{\bm X}]_{\rm ss})={\bm 0}$ can be cast into a form that allows for a closed-form graphical solution.
To this end, we introduce the Laplacian matrix $\mathsf{A}$ of the complex-reaction graph with off-diagonal elements
\begin{equation}
    A_{lm} = \begin{cases}
        k_{+\rho} & {\rm if} ~
        \bm{y}_l = \bm{\nu}_\rho^- ~ {\rm and} ~
        \bm{y}_m = \bm{\nu}_\rho^+, \\
        k_{-\rho} & {\rm if} ~
        \bm{y}_l = \bm{\nu}_\rho^+ ~ {\rm and} ~
        \bm{y}_m = \bm{\nu}_\rho^-, \\
        0 & {\rm otherwise},
    \end{cases}
\end{equation}
for $l \neq m$ and diagonal elements $A_{ll} = -\sum_{m\in\mathcal{C}\setminus\{l\}} A_{lm}$~\cite{gunawardena2003chemical,feinberg2019foundations}.
We also introduce the vector $\bm{\Psi}(\bm{x}) = \{ \bm{x}^{\bm{y}_l}\}_{l\in N_C}$ whose components are products of chemical species raised to a power determined by how often that species appears in a given complex.
Despite the additional complication of introducing these two quantities, it can be shown that the complex-balanced steady-state concentration now solves the alternative equation ${\mathsf C}\cdot {\bm J}([{\bm X}]_{\rm ss})=\mathsf{A} \cdot \bm{\Psi}([\bm{X}]_{\rm ss}) = {\bm 0}$.
The virtue of this reformulation is that the right null vectors of the Laplacian matrix can be given by a graphical solution based on the matrix-tree theorem~\cite{hill1966studies,gunawardena2012linear,owen2020universal}, which will also allow us to import some of the methodology from \cite{owen2020universal} to constrain the response.

\subsection{Linear CRNs ($\delta = 0$)}

\begin{figure}[t]
\centering
\includegraphics[width=0.75\columnwidth]{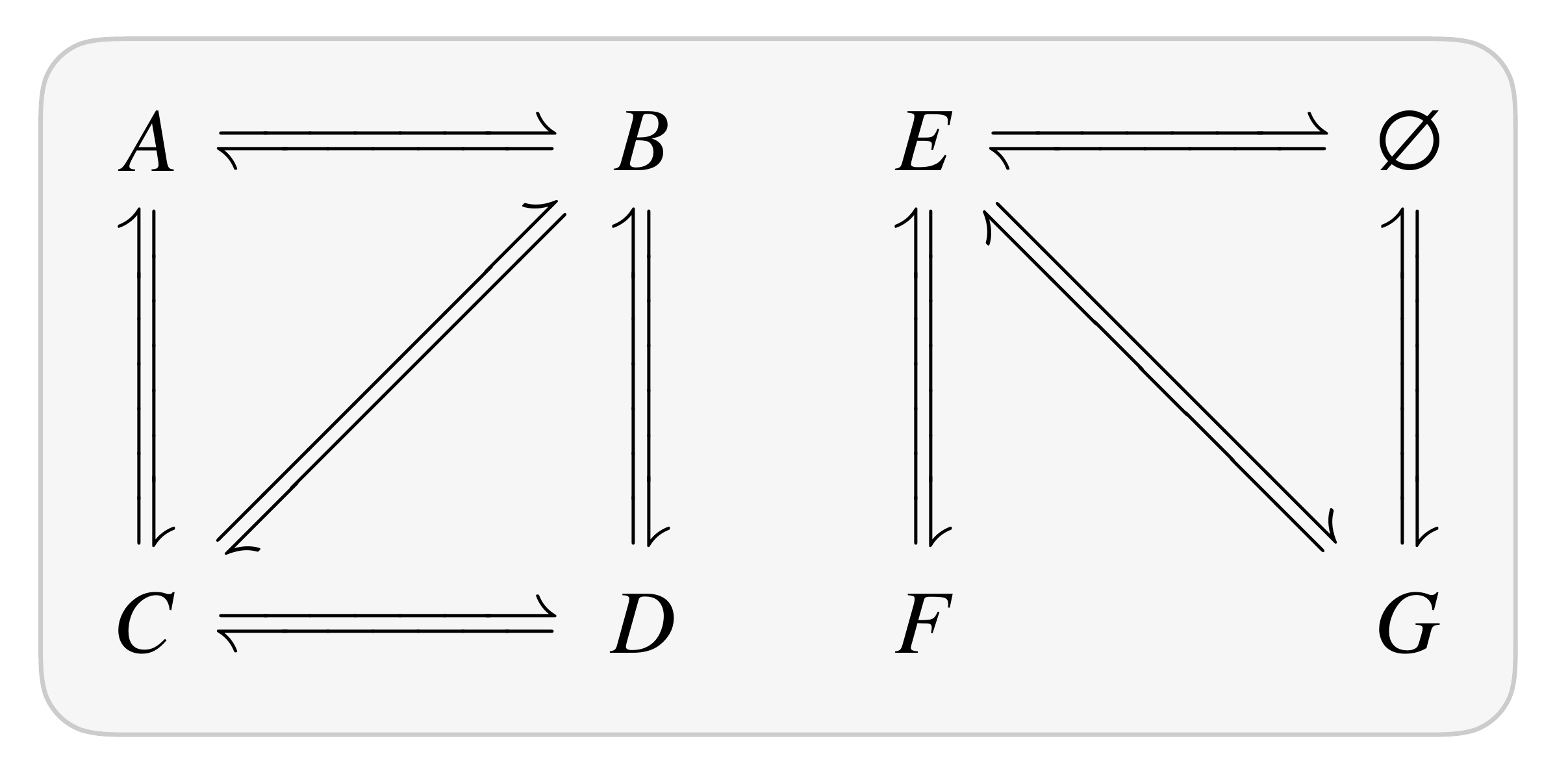}
\caption{Linear CRN with two linkage classes:
Each linkage class of a linear CRN can be regarded as an independent system.}
\label{fig:linear_CRN}
\end{figure}

Every linear CRN has deficiency zero.
Furthermore, in any  linear CRN, each linkage class can be regarded as an independent system, with chemical species in separate linkage classes evolving independently from each other (for example, $A$ and $E$ in Fig.~\ref{fig:linear_CRN}).
Thus, without loss of generality, for linear CRNs we need only consider a single linkage class.

For linear CRNs, the results of \cite{owen2020universal} allow us to directly bound the response by recognizing that the deterministic $\langle {\bm n}\rangle$ satisfies the same equation  $\mathsf{A} \cdot\langle {\bm n}\rangle ={\bm 0}$ as the steady-state distribution of a Markov jump process.  
For example, in a closed CRN with a single conservation law,  \cite{owen2020universal}  predicts that the single-rate perturbation is bounded by
\begin{equation}
\left |k_{\pm\rho}\frac{\partial\langle n_i\rangle}{\partial k_{\pm\rho}}\right|\le \langle n_i\rangle \left( 1 -\frac{\langle n_i\rangle}{n_{\rm tot}}\right)
\end{equation}
Connecting this upper bound to number fluctuations is possible, but requires information that the steady-state distribution is of product form~\eqref{eq:product_form}.
Below we take a different approach, building our analysis off the fluctuation-response equality~\eqref{eq:FRR}, which allows us to develop in parallel response bounds for linear and nonlinear deficiency zero networks.

To begin, we need to distinguish whether or not our linear CRN is open to exchange of chemical species with the surroundings, which would manifest as one of the complexes being the null vector.
If the null vector is a complex, e.g., as in the right linkage class in Fig.~\ref{fig:linear_CRN}, the CRN has no conservation law.
This can be deduced by noting the number of complexes is larger than the number of species $N_C = N_S + 1$ due to the presence of the null complex.
Furthermore, the number of independent reaction channels is simply one less than the number of complexes, ${\rm rank}({\mathsf S}^T)=N_C-1$, as each complex is linked to at least one reaction.
Then, by the rank-nullity theorem, there are no conservation laws: ${\rm dim}({\rm ker}({\mathsf S}^T))=N_S-{\rm rank}({\mathsf S}^T)=N_S-(N_C-1)=0$.
In this case, the steady-state distribution becomes a multivariate Poisson distribution~\cite{heuett2006grand,anderson2010product} over the entire microscopic state space, in which the numbers of different chemical species are uncorrelated, ${\rm Cov}\{n_i,n_j\} = {\rm Var}\{n_i\} \delta_{ij}$.
Then, the fluctuation-response relation \eqref{eq:FRR} can be written as
\begin{equation}\label{eq:FRR_lin_noncons}
    \frac{\partial \langle n_i \rangle}{\partial \lambda}
    = {\rm Var} \{n_i \}  \frac{\partial}{\partial \lambda} \ln [X_i]_{\rm ss}.
\end{equation}

On the other hand, if a linear CRN does not include a null vector as a complex,  then $N_C = N_S$ and there is a single conservation law, ${\rm dim}({\rm ker}({\mathsf S}^T))=N_S-{\rm rank}({\mathsf S}^T)=N_S-(N_C-1)=1$.
The one conservation law is the total number of chemical species $n_{\rm tot} = \sum_{i\in\mathcal{S}} n_i$.
In this case, the steady-state distribution becomes a multinomial distribution over a stoichiometric compatibility class with a single isolated connected component, whose mean and covariance are given by $\langle n_i \rangle = n_{\rm tot}[X_i]/\sum_{j\in\mathcal{S}}[X_j]$ and ${\rm Cov}\{n_i,n_j\} = \langle n_i \rangle ( \delta_{ij} - \langle n_j \rangle/n_{\rm tot})$~\cite{heuett2006grand,anderson2010product}.
This allows us to separate the sum in \eqref{eq:FRR} into two part with $j=i$ and $j\neq i$:
\begin{equation}
\begin{aligned}
    & \sum_{j\in\mathcal{S}\setminus\{i\}} {\rm Cov} \{n_i, n_j \}  \frac{\partial \ln [X_j]_{\rm ss}}{\partial \lambda} \\
    & = -{\rm Var}\{n_i\} \frac{\partial}{\partial \lambda} \ln \left( \sum_{j\in\mathcal{S}\setminus\{i\}} [X_j]_{\rm ss} \right).
\end{aligned}
\end{equation}
As a result, the fluctuation-response relation \eqref{eq:FRR} becomes
\begin{equation}\label{eq:FRR_lin_cons}
    \frac{\partial \langle n_i \rangle}{\partial \lambda}
    = {\rm Var}\{ n_i \} \frac{\partial}{\partial \lambda} \ln \left( \frac{[X_i]_{\rm ss}}{\sum_{j\in\mathcal{S}\setminus\{i\}} [X_j]_{\rm ss}} \right).
\end{equation}

To proceed, we recognize that for linear CRNs there is a one-to-one correspondence between chemical species and complexes.
Thus, we can always choose a basis for the complexes such that $\Psi_i([{\bm X}_{\rm ss}])=[X_i]_{\rm ss}$, and the last complex is the null vector $\Psi_{\bm \emptyset}([\bm{X}_{\rm ss}])=([{\bm X}]_{\rm ss})^{\bm 0}=1$ when it is present. 
In this way, the condition for the complex-balanced steady-state concentrations becomes $\mathsf{A}\cdot [\bm{X}]_{\rm ss} = \bm{0}$, where with a slight abuse of notation we write $[X_{N_S+1}]=1$ for the null vector.
In this way the steady-state concentrations  $[\bm{X}]_{\rm ss}$ have a graphical representation based on the matrix-tree theorem since $[\bm{X}]_{\rm ss}$ lies on the right null space of a Laplacian matrix $\mathsf{A}$~\cite{hill1966studies,gunawardena2012linear,owen2020universal}.
Such a graphical representation has been exploited recently to study the response of nonequilibrium Markov jump processes~\cite{owen2020universal}, whose master equation has the same structure.
The theory developed in \cite{owen2020universal} is directly applicable to the present setup.
The equation (F1) of \cite{owen2020universal} can be translated into
\begin{equation}\label{eq:inequality1}
    \left| k_{\pm \rho} \frac{\partial}{\partial k_{\pm \rho}} \ln \frac{[X_i]_{\rm ss}}{[X_j]_{\rm ss}} \right|
    \leq 1
\end{equation}
and
\begin{equation}
    \left| k_{\pm \rho} \frac{\partial}{\partial k_{\pm \rho}} \ln \left( \frac{[X_i]_{\rm ss}}{\sum_{j\in\mathcal{S}\setminus\{i\}} [X_i]_{\rm ss}} \right) \right|
    \leq 1.
\end{equation}
Although the counterpart of the former inequality does not appear in \cite{owen2020universal} explicitly, it is straightforwardly obtainable by applying the same logic to prove (F1) of \cite{owen2020universal}.
Similarly, the equations (C12) and (C13) of \cite{owen2020universal} can be translated respectively into
\begin{equation}\label{eq:inequality3}
    \left| \frac{\partial}{\partial B_\rho}  \ln \frac{[X_i]_{\rm ss}}{[X_j]_{\rm ss}} \right|
    \leq \tanh\left( \frac{\mathcal{F}}{4} \right)
\end{equation}
and
\begin{equation}\label{eq:inequality4}
    \left| \frac{\partial}{\partial B_\rho} \ln \left( \frac{[X_i]_{\rm ss}}{\sum_{j\in\mathcal{S}\setminus\{i\}} [X_j]_{\rm ss}} \right) \right|
    \leq \tanh\left( \frac{\mathcal{F}}{4} \right),
\end{equation}
Note that only chemical cycles visible in the complex-reaction graph are taken into account for defining $\mathcal{F}$ since the matrix-tree theorem only uses information encoded in $\mathsf{A}$, i.e., the connectivity between complexes.
The zero deficiency then ensures all chemical cycles are visible in the complex-reaction graph.

To complete our analysis, we apply inequalities (\ref{eq:inequality1})-(\ref{eq:inequality4}) to \eqref{eq:FRR_lin_noncons} and \eqref{eq:FRR_lin_cons}, to derive trade-offs for linear CRNs 
\begin{equation}
    \left| k_{\pm \rho} \frac{\partial \langle n_i \rangle}{\partial k_{\pm \rho}} \right| \leq {\rm Var}\{n_i \}
\end{equation}
and
\begin{equation}
    \left| \frac{\partial \langle n_i \rangle}{\partial B_{\rho}} \right| \leq {\rm Var}\{n_i\} \tanh\left(\frac{\mathcal{F}}{4}\right),
\end{equation}
which hold regardless of the conservation law.

This analysis complements a derivation based directly on the results of \cite{owen2020universal}.

\subsection{Nonlinear deficiency zero ($\delta = 0$): Theory}

The explicit form of the mean and covariance are generally unavailable for nonlinear CRNs.
In addition, different linkage classes are generally not independent.
These facts significantly complicate the analysis, requiring us to investigate the response of the full steady-state distribution.

While we are not able to prove response is bounded by number fluctuations, we are able to show it is limited by another measure of fluctuations in chemical space that we now introduce. 
Details can be found in Appendix~\ref{sec:deficiency_zero_nonlinear_CRNs}.
To this end, we define a matrix formed by keeping only the linearly independent columns of the stoichiometric matrix
\begin{equation}\label{eq:reduced_stoichiometric_matrix}
    \mathsf{S}' = \begin{pmatrix}
    \vert & \vert & & \vert \\
    \Delta\bm{\nu}_1 & \Delta\bm{\nu}_2 & \cdots & \Delta\bm{\nu}_{s} \\
    \vert & \vert & & \vert
    \end{pmatrix}
\end{equation}
Since all the columns are linearly independent, $\mathsf{S}'$ has a left inverse matrix $(\mathsf{S}')^{-1}$ satisfying $(\mathsf{S}')^{-1} \mathsf{S}'= \mathsf{I}$.
It is worth mentioning that ${\mathsf S}'$ is not unique, in that we could have chosen a different subset of linearly independent reactions to form ${\mathsf S}'$.

Applying the matrix tree theorem to the response of the log-ratio of the steady-state probabilities of two states, we obtain the bounds 
\begin{equation}\label{eq:deficiency_zero_ineq1}
    \left| k_{\pm\rho} \frac{\partial \langle n_i \rangle_\Gamma}{\partial k_{\pm\rho}} \right|
    \leq \sum_{\sigma\in\mathcal{R}'} \left|  \sum_{j\in\mathcal{S}} (\mathsf{S}')^{-1}_{\sigma j} {\rm Cov}_\Gamma \{ n_i, n_j \} \right|
\end{equation}
and
\begin{equation}\label{eq:deficiency_zero_ineq2}
    \left| \frac{\partial \langle n_i \rangle_\Gamma}{\partial B_\rho} \right|
    \leq \sum_{\sigma\in\mathcal{R}'}  \left| \sum_{j\in\mathcal{S}}(\mathsf{S}')^{-1}_{\sigma j} {\rm Cov}_\Gamma \{ n_i, n_j \} \right| \tanh\left(\frac{\mathcal{F}}{4}\right).
\end{equation}
where $\mathcal{R}' = \{ 1, 2, \cdots, s \}$.  
These weighted covariances depend on tangible observable number fluctuations though are coupled  through the network topology.
Further insight into the nature of these fluctuations is offered by considering another representation of the population vectors ${\bm{n}}$, detailed in Appendix~\ref{sec:deficiency_zero_nonlinear_CRNs}.
Any state vector $\bm{n}'$ can be reached from any other state vector $\bm{n}$ through a unique series of the  $s$ independent reactions in the columns of  $\mathsf{S}'$ \eqref{eq:reduced_stoichiometric_matrix}.
If we denote $\mu_\sigma({\bm n}, {\bm n}')$ as the number of times reaction $\sigma$ occurs along the reaction path from ${\bm n}$ to ${\bm n}'$, then the weighted covariances can be re-expressed as 
\begin{equation}
     \sum_{j \in \mathcal{S}} (\mathsf{S}')^{-1}_{\sigma j} {\rm Cov}_\Gamma \{ n_i, n_j \} = \sum_{\bm{n},\bm{n}'\in\Gamma}
    n_i \mu_\sigma(\bm{n},\bm{n}')
    \pi_\Gamma(\bm{n}) \pi_\Gamma(\bm{n}'),
\end{equation}
in terms of a correlation depending on the distance between the states ${\bm n}$ and ${\bm n}'$ measured in reactions.
In this way, response can be viewed as bounded by number fluctuations or fluctuations in reaction distance.

Although we were not able to derive the compact forms of trade-offs \eqref{eq:main1} and \eqref{eq:main2}, numerical evidence presented in the next subsection suggests they may be valid for a broad class of nonlinear CRNs as well.

\subsection{Nonlinear deficiency zero ($\delta = 0$): Numerics}\label{sec:deficiency_zero_numerical}

We provide numerical results for several nonlinear, deficiency zero CRNs to demonstrate the potential validity of the trade-offs in \eqref{eq:main1} and \eqref{eq:main2}, allowing for multiple conservation laws and multiple chemical cycles.  
Without loss of generality, we consider here only one isolated connected component.
This follows because the total response $\partial \langle n_i \rangle/\partial \lambda$ is a linear combination of the partial responses $\partial \langle n_i \rangle_\Gamma /\partial \lambda$.
If the trade-offs \eqref{eq:main1} and \eqref{eq:main2} hold for each partial response, then the total response is also bounded by number fluctuations,
\begin{equation}
\left| \frac{\partial\langle n_i\rangle}{\partial \lambda}\right|\le \sum_\Gamma p_\Gamma {\rm Var}_\Gamma\{n_i\} \le  {\rm Var}\{n_i \}
\end{equation} 
due to the convexity of the variance.
We find in each case studied here that trade-offs  \eqref{eq:main1} and \eqref{eq:main2} hold with common properties, $\alpha_1 = \alpha_2$ and $g(\mathcal{F}) = \tanh(\mathcal{F}/4)$.
The values of $\alpha_1 = \alpha_2$ depend on the chemical species of interest and the perturbed chemical reaction but not on the size of the stoichiometric compatibility class.
 Details of the numerical methods can be found in Appendix~\ref{sec:diffusion_approximation}.

In our numerical analysis, we will distinguish two methodologies for numerically determining the response and fluctuations based on system size.
When the size of the stoichiometric compatibility class is small enough, say due to the existence of conservation laws, such that the Chemical Master Equation \eqref{eq:CME1} is practically solvable, we will use it to directly calculate the steady-state distribution as well as the steady-state response.  
Then to examine the trade-offs \eqref{eq:main1} and \eqref{eq:main2}, we will plot two quality factors
\begin{align}
{\mathcal K}(\pm\rho,X_i) &= \frac{1}{{\rm Var}\{n_i\}} \left|k_{\pm\rho}\frac{\partial\langle n _i\rangle}{\partial k_{\pm\rho}}\right|,\\
{\mathcal B}(\rho,X_i) &= \frac{1}{{\rm Var}\{n_i\}} \left|\frac{\partial\langle n _i\rangle}{\partial B_{\rho}}\right|.
\end{align}
Other times it will be more feasible to determine the response and fluctuations in the macroscopic limit, using the Chemical Rate Equation \eqref{eq:CRE1} to determine the steady-state concentration, and the Linear Noise Approximation to estimate the fluctuations.
To distinguish this case, we introduce modified notation for the quality factors
\begin{align}
{\mathcal K}^{\rm M}(\pm\rho,X_i) &= \frac{1}{D_i} \left|k_{\pm\rho}\frac{\partial[X_i]_{\rm ss}}{\partial k_{\pm\rho}}\right|, \\
{\mathcal B}^{\rm M}(\rho,X_i) & = \frac{1}{D_i} \left|\frac{\partial[X_i]_{\rm ss}}{\partial B_{\rho}}\right|,
\end{align}
where the superscript `M' signifies macroscopic.
Recalling that the Chemical Rate Equation can have multiple stable fixed points, we avoid this complication by only considering the macroscopic limit in situations where there is a unique stable solution.

\subsubsection{Single chemical cycle and single conservation law $(\delta = 0)$}

\begin{figure}[t]
\centering
\includegraphics[width=0.9\columnwidth]{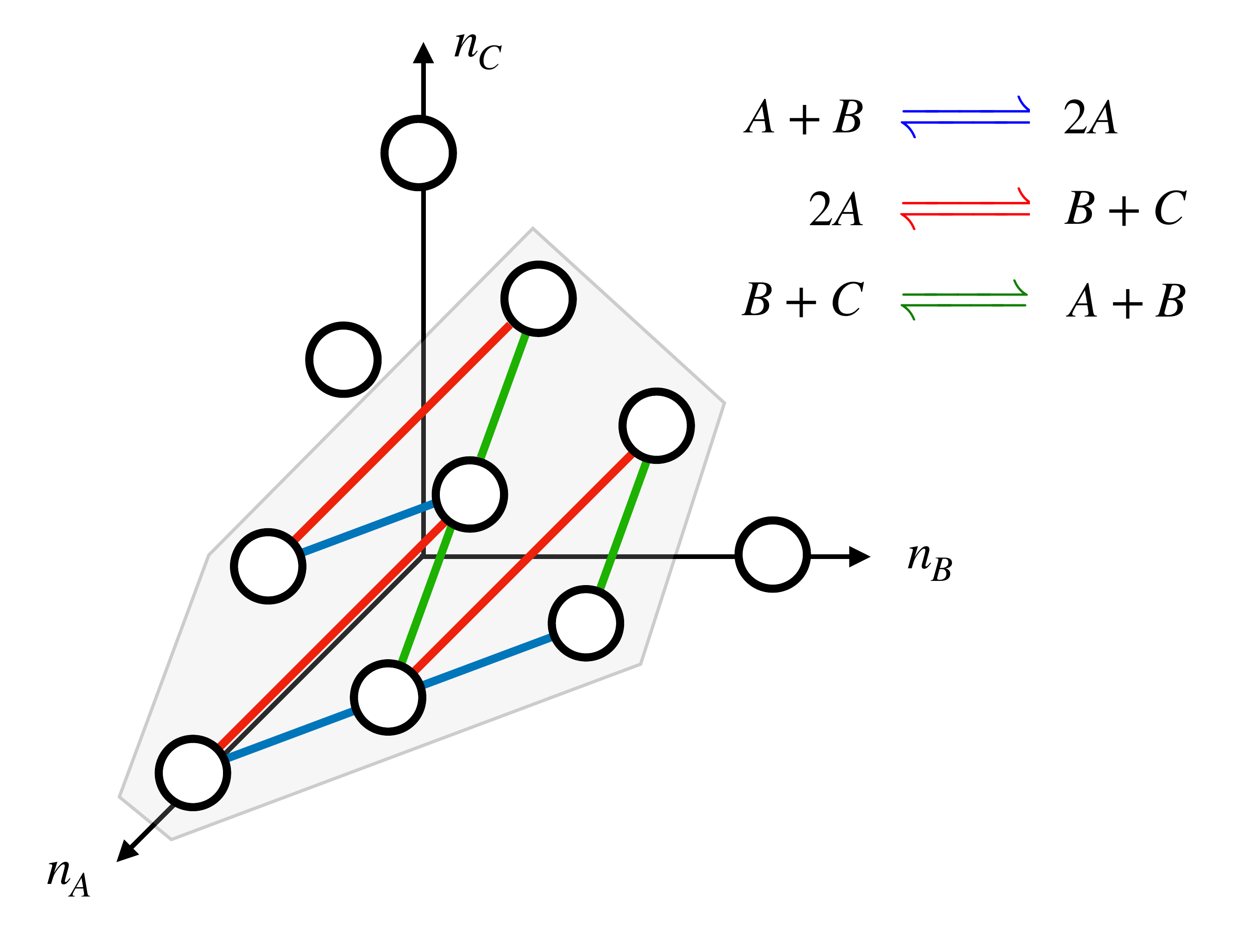}
\caption{
Stoichiometric compatibility class of chemical reaction network \eqref{eq:ABC_model_deficiency_0} with total number $\Lambda = n_A + n_B + n_C = 3$.
It consists of four isolated connected components, including three isolated states $(0,3,0)$, $(0,0,3)$, and $(1,0,2)$.
The shaded area highlights the largest isolated connected component.}
\label{fig:state_space_deficiency_zero}
\end{figure}

We first consider a deficiency zero, nonlinear CRN with a single chemical cycle and a single conservation law, with complex-reaction graph
\begin{equation}\label{eq:ABC_model_deficiency_0}
\schemestart
    \chemfig{2A} \arrow{<=>[\tiny $k_{+2}$][\tiny $k_{-2}$]} \chemfig{B+C}
    \arrow{<=>[\tiny $k_{-3}$][\tiny $k_{+3}$]}[125] \chemfig{A+B}
    \arrow{<=>[\tiny $k_{-1}$][\tiny $k_{+1}$]}[235]
\schemestop.
\end{equation}
This CRN is constructed from a slight modification of the example depicted in Fig.~\ref{fig:example_CRN} to make it deficiency zero, but with the same  stoichiometric matrix and conservation law.
The total number of chemical species $\Lambda = n_A + n_B + n_C$ is conserved, and the only chemical cycle is visible in the complex-reaction graph.
For fixed $\Lambda$ the stoichiometric compatibility class consists of a large isolated connected component  and three isolated states, $(n_A,n_B,n_C) \in \{ (0, \Lambda, 0), ~ ( 0, 0, \Lambda), ~ (1, 0, \Lambda-1) \}$.
The case of $\Lambda=3$ is depicted in Fig.~\ref{fig:state_space_deficiency_zero}.
The steady-state probabilities of the three isolated states are fixed by the initial distribution and do not change when the dynamics are perturbed.
Accordingly, it is sufficient to consider only the large isolated connected component to examine the trade-offs since the isolated states do not contribute to response and number fluctuations.

By choosing a basis where $\{ X_1 = A, X_2 = B, X_3 = C \}$ and the two independent chemical reactions $\{ A+B \rightleftharpoons 2A, 2A \rightleftharpoons B+C \}$, we have the reduced stoichiometric matrix 
\begin{equation}
    \mathsf{S}' = \begin{pmatrix}
    1 & -2 \\
    -1 & 1 \\
    0 & 1
    \end{pmatrix}
\end{equation}
and its left inverse 
\begin{equation}
    (\mathsf{S}')^{-1} = \begin{pmatrix}
    0 & -1 & 1 \\
    -1/3 & -1/3 & 2/3
    \end{pmatrix}.
\end{equation}
Plugging the components of $(\mathsf{S}')^{-1}$ into \eqref{eq:deficiency_zero_ineq1} and \eqref{eq:deficiency_zero_ineq2} and using ${\rm Cov}\{n_X, \Lambda\} = 0$, we have inequalities
\begin{equation}\label{eq:ineq_ABC1}
\begin{aligned}
    \left| k_{\pm\rho} \frac{\partial \langle n_X \rangle}{\partial k_{\pm\rho}} \right|
    & \leq | {\rm Cov}\{n_X, n_B - n_C\} | \\
    & ~~~ + | {\rm Cov}\{n_X,n_C\} |
\end{aligned}
\end{equation}
and
\begin{equation}\label{eq:ineq_ABC2}
\begin{aligned}
    \left| \frac{\partial \langle n_X \rangle}{\partial B_{\rho}} \right|
    & \leq ( | {\rm Cov}\{n_X, n_B - n_C \} | \\
    & ~~~ + | {\rm Cov}\{n_X,n_C\} | )
    \tanh\left(\frac{\mathcal{F}}{4}\right),
\end{aligned}
\end{equation}
where $X \in \{ A, B, C\}$ and $\mathcal{F}$ is the absolute value of the thermodynamic force associated with the sole chemical cycle \eqref{eq:maxF}.

\begin{figure}[t]
\centering
\includegraphics[width=\columnwidth]{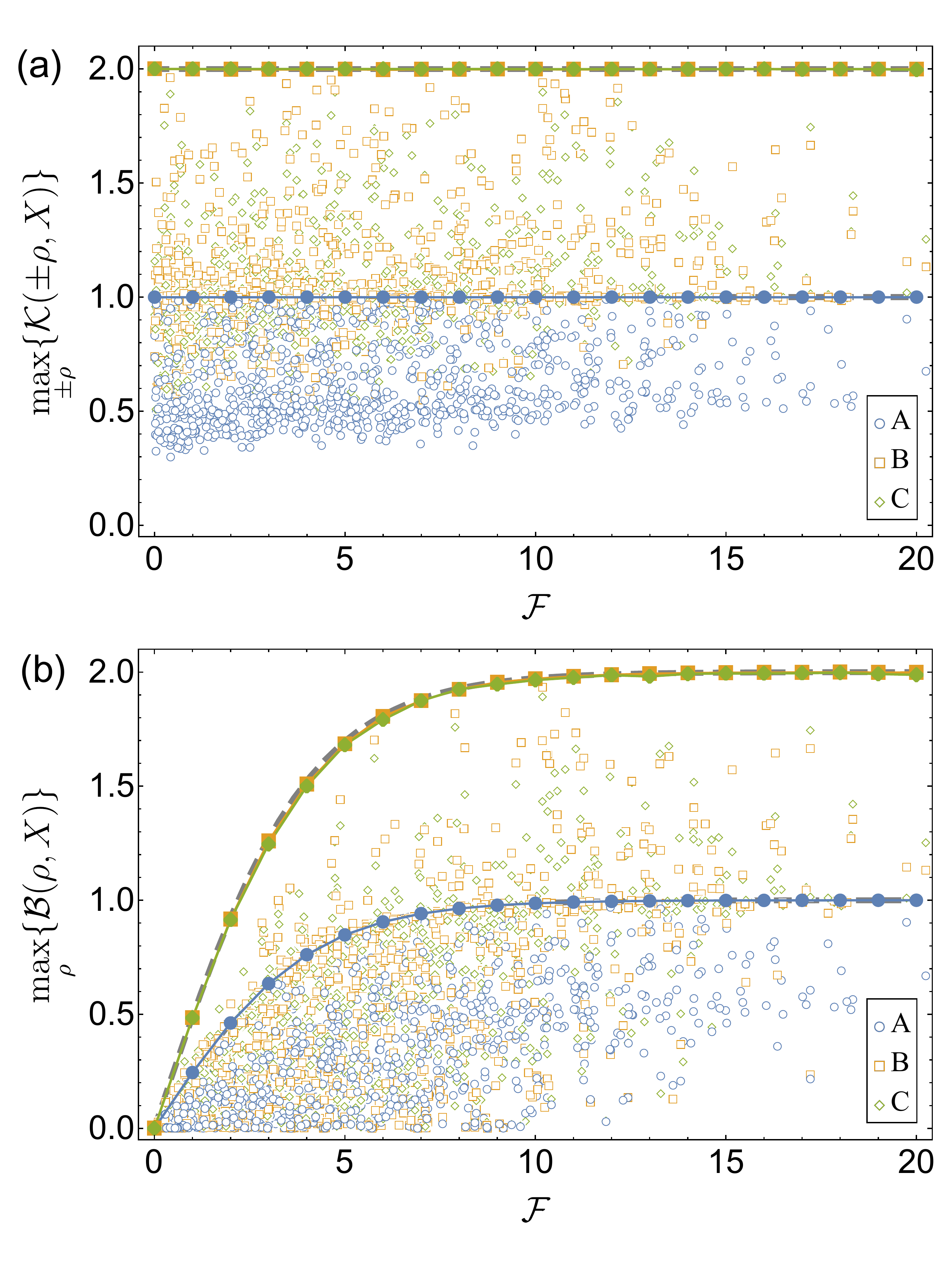}
\caption{The maximum quality factors among different choices of $\rho$ for the model in \eqref{eq:ABC_model_deficiency_0} at a fixed $(X,\Lambda, \{ k_{\pm \rho}\}_{\rho=1}^3, \Omega)$.
Each open symbol is obtained from a set of randomly sampled rate constants $\{ k_{\pm \rho}\}_{\rho=1}^3$ and system size $\Omega$, the logs of which are uniformly sampled from the range $[-5,5]$, with $\Lambda$ sampled from  $\{3,4,5,6,7\}$.
The number of open symbols at a fixed $(X, \Lambda)$ in each panel is $200$.
The filled symbols are the maximum values at a fixed $\mathcal{F}$ and chemical species, which are numerically optimized using simulated annealing.
The perturbed edges used for the optimization are $A+B\rightharpoonup2A$ for (a) and $A+B\rightleftharpoons 2A$ for (b), and $\Lambda$ is fixed by 3.
The dashed lines represent $1$ and $2$ in (a), $\tanh(\mathcal{F}/4)$ and $2\tanh(\mathcal{F}/4)$ in (b).}
\label{fig:deficiency_zero_ABC}
\end{figure}

We numerically calculated the two quality factors ${\mathcal K}(\pm\rho,X)$ and ${\mathcal B}(\rho,X)$ for each species $X$ and reaction channel $\rho$ for random sets of system parameters $(\Lambda, \{ k_{\pm \rho}\}_{\rho=1}^3, \Omega)$.
Then for each random set of system parameters, we plot in Fig.~\ref{fig:deficiency_zero_ABC} the maximum quality factors for each species $X$, maximized over perturbed reaction channels, depicted as open symbols.
The figure suggests the existence of upper bounds on the response that do not appear to depend on $\Lambda$.
To confirm this observation and determine the shape of the upper bound, we numerically optimized the quality factors using simulated annealing~\cite{kirkpatrick1983optimization}.
The results from the optimization are drawn as filled symbols in Fig.~\ref{fig:deficiency_zero_ABC}, which align with the trade-offs \eqref{eq:main1} and \eqref{eq:main2} with $g(\mathcal{F})=\tanh(\mathcal{F}/4)$ regardless of $(\rho, X)$.
A closer investigation further reveals that the two prefactors $\alpha_1$ and $\alpha_2$ coincide for each pair $(\rho, X)$ (see TABLE~\ref{tbl:ABC_model_deficiency_0}).

\begin{table}[t]
\centering
\begin{tabular}{cccc}
\toprule
\multicolumn{1}{c}{} & \multicolumn{3}{c}{\textbf{chemical reaction}} ($\rho$) \\
\cmidrule(rl){2-4} 
\textbf{chemical species} ($X$) & ~ $\pm 1$ & ~~ $\pm 2$ & ~~ $\pm 3$ \\
\hline\hline
$A$ & ~ (1,~1) & ~~ (1,~1) & ~~ (1,~1) \\
$B$ & ~ (2,~2) & ~~ (1,~1) & ~~ (2,~2) \\
$C$ & ~ (2,~2) & ~~ (1,~1) & ~~ (2,~2) \\
\bottomrule
\end{tabular}
\caption{Numerically estimated values of ($\alpha_1$, $\alpha_2$) for each pair $(\rho,X)$ in \eqref{eq:ABC_model_deficiency_0}.}
\label{tbl:ABC_model_deficiency_0}
\end{table}

\begin{figure}[t]
\centering
\includegraphics[width=\columnwidth]{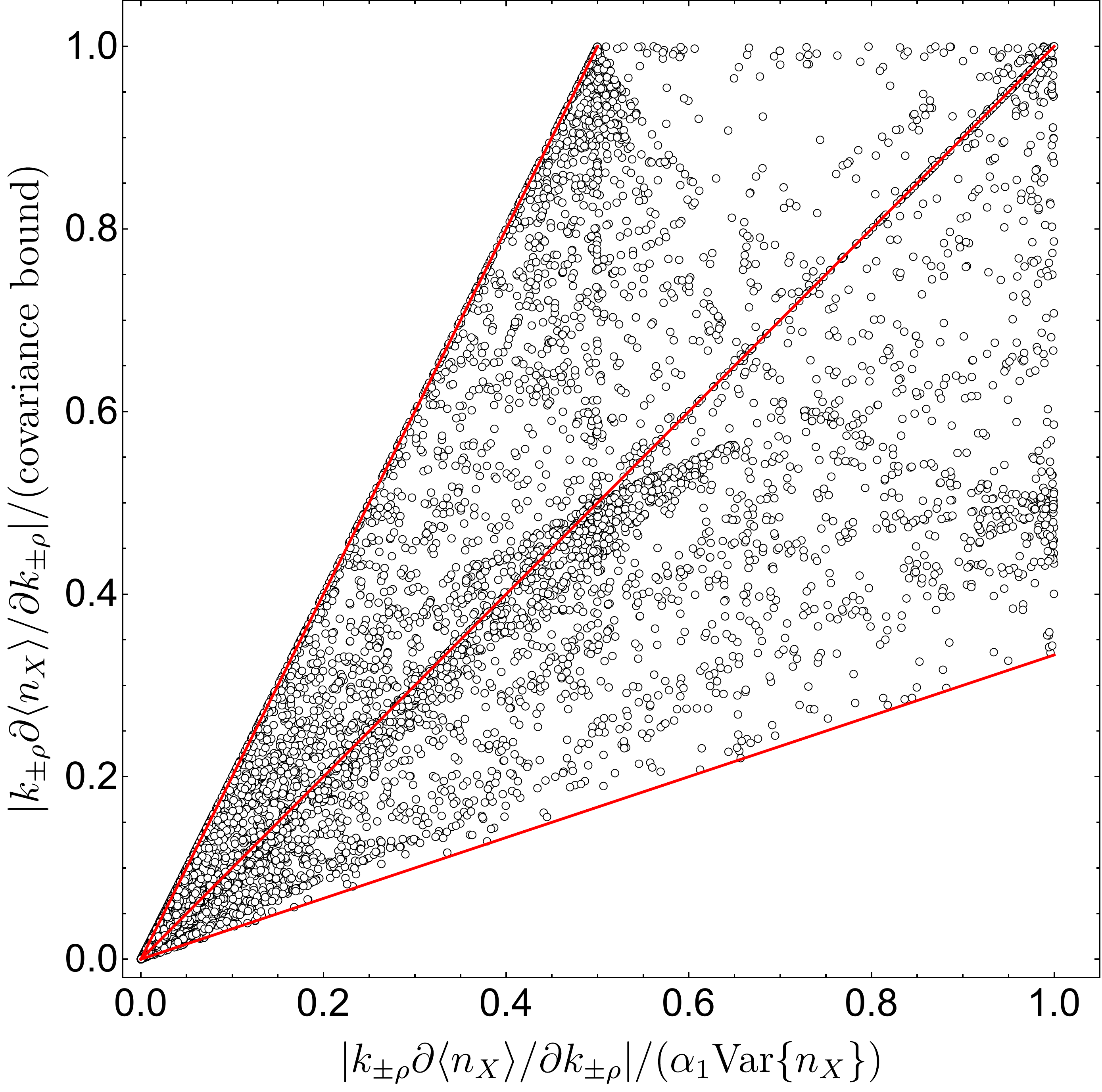}
\caption{Comparison of the tightness of the proven bound \eqref{eq:ineq_ABC1} and conjectured bound \eqref{eq:main1} for the model in \eqref{eq:ABC_model_deficiency_0}.
The covariance bound in the vertical axis refers to the right-hand side of \eqref{eq:ineq_ABC1}.
Each symbol is obtained from a set of randomly sampled rate constants $\{ k_{\pm \rho}\}_{\rho=1}^3$ and system size $\Omega$, the logs of which are uniformly sampled from the range $[-5,5]$, with $\Lambda = 3$.
The coefficient $\alpha_1$ is used as the estimated values in TABLE~\ref{tbl:ABC_model_deficiency_0}.
The number of symbols for a fixed pair of $(X, \pm \rho)$ is $500$.
The red guidelines have slopes of $1/3$, $1$, and $2$.}
\label{fig:comparison_ABC}
\end{figure}

To investigate the relationship between our covariance bounds \eqref{eq:ineq_ABC1} and the compact trade-offs \eqref{eq:main1}, we compare the tightness of the bounds in Fig.~\ref{fig:comparison_ABC} by plotting the ratios of the response to fluctuations for single rate perturbations; similar conclusions would be reached by considering the force-dependent symmetric perturbations bounded in \eqref{eq:ineq_ABC2} and \eqref{eq:main2}.
The numerical results validate both inequalities as all samples are confined to the unit square.  
Moreover we see that one bound is generally not tighter than the other as samples fall above and below the diagonal, implying that the covariance bounds \eqref{eq:ineq_ABC1} and \eqref{eq:ineq_ABC2} we have derived are not sufficiently tight in this case to prove the compact trade-offs \eqref{eq:main1} and \eqref{eq:main2}.
Nevertheless, the enveloping guidelines shown in red based on the inequalities, $(\alpha_1/2) {\rm Var}\{ n_X \} \leq |{\rm Cov}\{n_X, n_B - n_C\}| + |{\rm Cov}\{ n_X, n_C \}| \leq 3\alpha_1 {\rm Var} \{ n_ X \}$, imply the bounds are not completely independent.

\subsubsection{Multiple conservation laws $(\delta = 0)$}

\begin{figure}[t]
\centering
\includegraphics[width=\columnwidth]{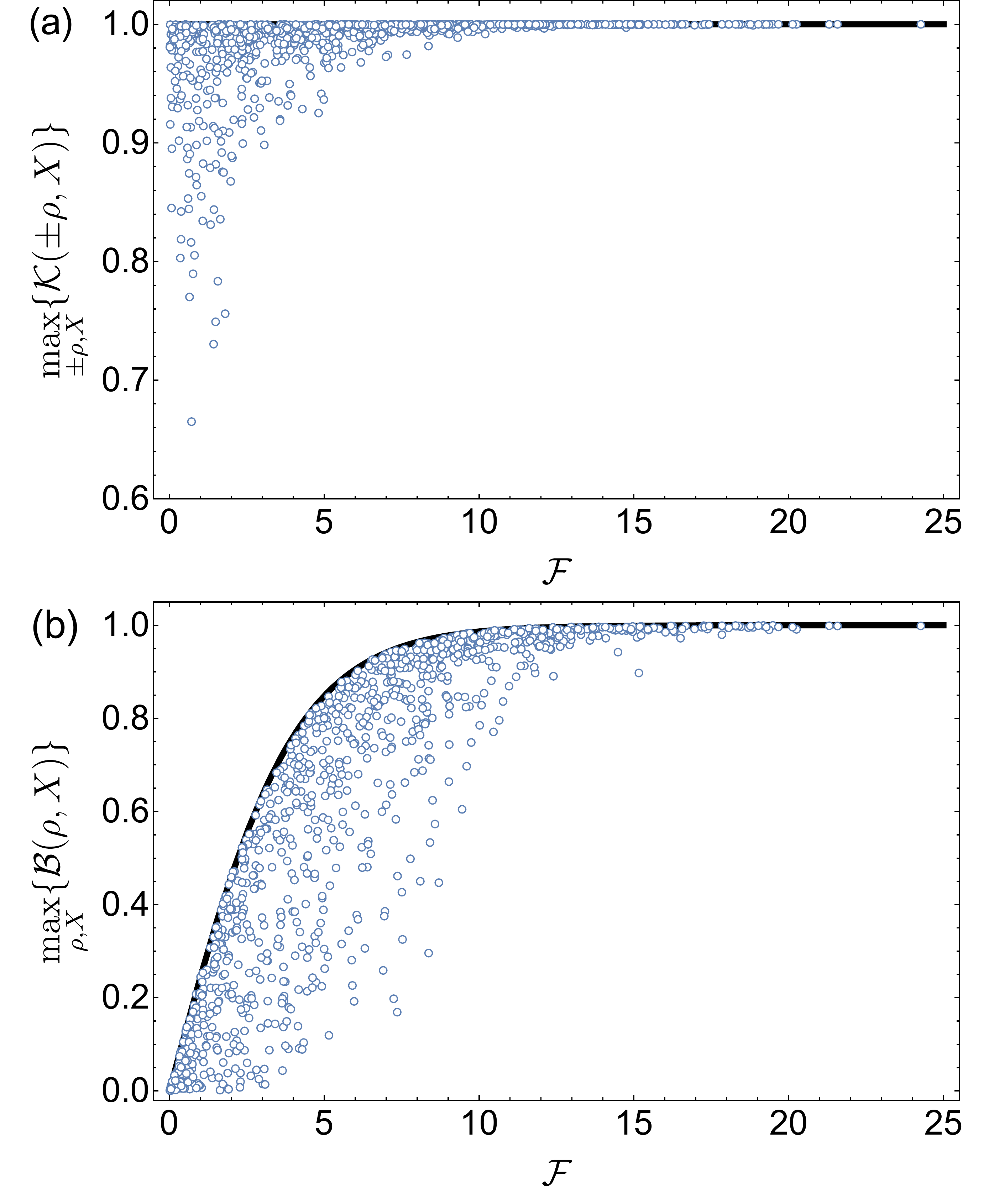}
\caption{
The maximum quality factors among different choices of ($\rho$, $X$) for the model in \eqref{eq:modified_PdPC} at a fixed $(\Lambda_E, \Lambda_S, \{ k_{\pm \rho}\}_{\rho=1}^4, \Omega)$.
Each symbol is obtained from a set of randomly sampled rate constants $\{ k_{\pm \rho}\}_{\rho=1}^4$ and system size $\Omega$, the logs of which are uniformly sampled from $[-5,5]$.
The total number of enzymes and substrates, $(\Lambda_E, \Lambda_S)$, take values from $\{(1,5),(2,4),(3,3),(4,2),(5,1)\}$.
The number of symbols at a fixed $(\Lambda_E, \Lambda_S)$ in each panel is $200$.
The thick lines represent $1$ in (a), and $\tanh(\mathcal{F}/4)$ in (b).
}
\label{fig:modified_PdPC}
\end{figure}

To see whether and how the number of conservation laws affects the trade-offs, we consider a CRN with multiple conservation laws, but a single chemical cycle.  
Our model is composed of a substrate that can exist in an unmodified $S$ and modified form $P$.
In cells, this modification is often the addition of a phosphate group, but we will not make that specification here.
The modification and demodification are carried out by a single enzyme $E$ through two distinct intermediate species $ES_1$ and $ES_2$, resulting in the complex-reaction graph
\begin{equation}\label{eq:modified_PdPC}
\schemestart
    \chemfig{ES_2} \arrow{<=>[\tiny $k_{-3}$][\tiny $k_{+3}$]} \chemfig{E+P}
    \arrow{<=>[\tiny $k_{-2}$][\tiny $k_{+2}$]}[90] \chemfig{ES_1}
    \arrow{<=>[\tiny $k_{+1}$][\tiny $k_{-1}$]}[180] \chemfig{E + S}
    \arrow{<=>[\tiny $k_{-4}$][\tiny $k_{+4}$]}[270]
\schemestop.
\end{equation}
This CRN was inspired by the Goldbeter-Koshland model for the phosphorylation-dephosphorylation cycle (PdPC), which is a known mechanism for ultra-sensitivity.
The model was modified to make it reversible and deficiency zero.
Later in Sec.~\ref{sec:GB}, we will analyze the full (reversible) Goldbeter-Koshland model.

The modification-demodification cycle in \eqref{eq:modified_PdPC} has two independent conservation laws: the total numbers of enzymes $\Lambda_E = n_E + n_{ES_1} + n_{ES_2}$ and substrates $\Lambda_S = n_S + n_P + n_{ES_1} + n_{ES_2}$.
By choosing a basis where $\{ X_1 = E, X_2 = S, X_3 = P, X_4 = ES_1, X_5 = ES_2 \}$ and three independent reaction channels $\{ E+S \rightleftharpoons ES_1, ES_1 \rightleftharpoons E+P, E+P \rightleftharpoons ES_2 \}$, we have the reduced stoichiometric matrix 
\begin{equation}
    \mathsf{S}' = \begin{pmatrix}
    -1 & 1 & -1 \\
    -1 & 0 & 0 \\
    0 & 1 & -1 \\
    1 & -1 & 0 \\
    0 & 0 & 1
    \end{pmatrix}
\end{equation}
and its left inverse
\begin{equation}
    (\mathsf{S}')^{-1} = \begin{pmatrix}
    -1/4 & -5/8 & 3/8 & 1/8 & 1/8 \\
    0 & -1/2 & 1/2 & -1/2 & 1/2 \\
    -1/4 & -1/8 & -1/8 & -3/8 & 5/8
    \end{pmatrix}.
\end{equation}
Plugging the components of $(\mathsf{S}')^{-1}$ into \eqref{eq:deficiency_zero_ineq1} and \eqref{eq:deficiency_zero_ineq2} and using ${\rm Cov}\{n_X, \Lambda_E\} = {\rm Cov}\{n_X, \Lambda_S\} = 0$, we have the inequalities
\begin{equation}\label{eq:ineq_modified_PdPC}
\begin{aligned}
    \left| k_{\pm\rho} \frac{\partial \langle n_X \rangle}{\partial k_{\pm\rho}} \right|
    & \leq | {\rm Cov}\{n_X,n_S\} | + | {\rm Cov}\{n_X,n_{ES_2}\} | \\
    & ~~~ + | {\rm Cov}\{n_X,n_P + n_{ES_2}\} |
\end{aligned}
\end{equation}
and
\begin{equation}\label{eq:ineq_modified_PdPC2}
\begin{aligned}
    \left| \frac{\partial \langle n_X \rangle}{\partial B_{\rho}} \right|
    & \leq ( | {\rm Cov}\{n_X,n_S\} | + | {\rm Cov}\{n_X,n_{ES_2} \} | \\
    & ~~~ + | {\rm Cov}\{n_X,n_P + n_{ES_2}\} | ) \tanh\left(\frac{\mathcal{F}}{4}\right),
\end{aligned}
\end{equation}
where $X \in \{ E, S, P, ES_1, ES_2 \}$ and $\mathcal{F}$ is the absolute value of the thermodynamic force associated with the only chemical cycle \eqref{eq:maxF}.

We numerically calculated the maximum values ${\mathcal K}(\pm\rho,X)$ and ${\mathcal B}(\rho,X)$ among different pairs of ($\rho$, $X$) at fixed system parameters $(\Lambda_E, \Lambda_S, \{ k_{\pm \rho}\}_{\rho=1}^4, \Omega)$.
Rate constants $\{ k_{\pm \rho} \}_{\rho=1}^{4}$ and system size $\Omega$ are randomly chosen.
The total numbers of enzymes and substrates are selected from $(\Lambda_E, \Lambda_S) \in \{ (1,5), ~(2,4), ~(3,3), ~(4,2), ~(5,1) \}$.
The results are shown in Fig.~\ref{fig:modified_PdPC}, which demonstrate that the maximum quality factors are bounded from above by ${\mathcal K}\le 1$ and ${\mathcal B}\le \tanh(\mathcal{F}/4)$.
A detailed investigation further shows that the model \eqref{eq:modified_PdPC} obeys trade-offs, $|k_{\pm}\partial\langle n_X \rangle/\partial k_{\pm\rho}| \leq 1$ and $|\partial\langle n_X \rangle/\partial B_{\rho}| \leq \tanh(\mathcal{F}/4)$ no matter the choice of pair $(\rho, X)$.

\begin{figure}[t]
\centering
\includegraphics[width=\columnwidth]{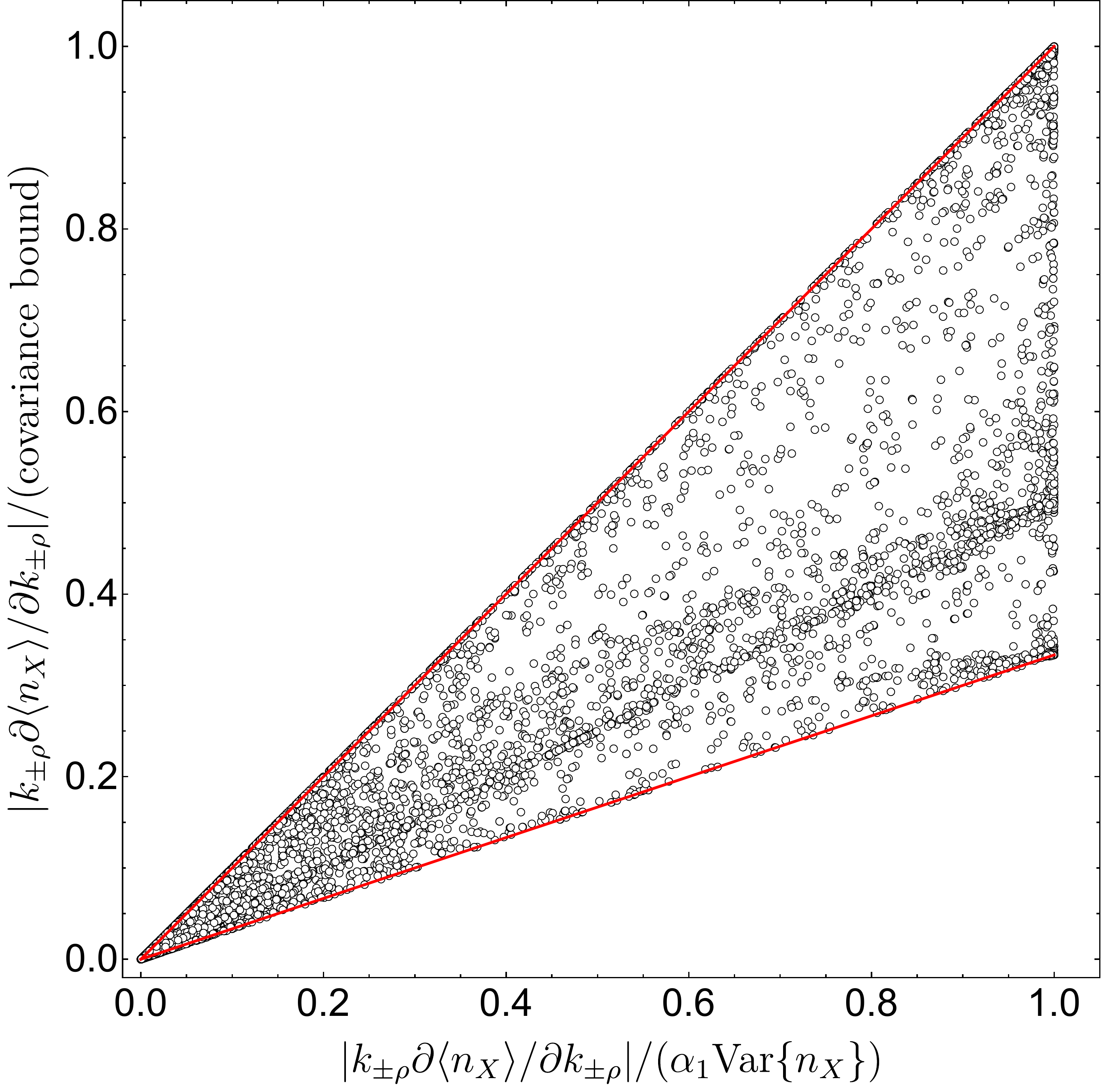}
\caption{Comparison of the tightness of the proven bound \eqref{eq:ineq_modified_PdPC} and conjectured bound \eqref{eq:main1} for the model in \eqref{eq:modified_PdPC}.
The covariance bound in the vertical axis refers to the right-hand side of \eqref{eq:ineq_modified_PdPC}.
Each symbol is obtained from a set of randomly sampled rate constants $\{ k_{\pm \rho}\}_{\rho=1}^4$ and system size $\Omega$, the logs of which are uniformly sampled from the range $[-5,5]$, with $(\Lambda_E, \Lambda_S) = (3,3)$.
The coefficient $\alpha_1$ is set to $1$ for all pairs of $(X, \pm \rho)$.
The number of symbols for a fixed pair of $(X, \pm \rho)$ is $200$.
The guidelines have slopes of $1/3$ and $1$.}
\label{fig:comparison_PdPC}
\end{figure}

In Fig.~\ref{fig:comparison_PdPC}, we numerically compare the covariance bound \eqref{eq:ineq_modified_PdPC} with the compact trade-offs \eqref{eq:main1}  for a single rate perturbation by plotting the ratio of response to the fluctuations.
The validity of both bonds is apparent as all samples fall inside the unit square.
Here, however, all samples fall below the diagonal implying that covariance bound \eqref{eq:ineq_modified_PdPC}  is tighter then the compact trade-off \eqref{eq:main1} in contrast to what was observed for the model in \eqref{eq:ABC_model_deficiency_0}.
Together the results in Figs.~\ref{fig:comparison_ABC} and \ref{fig:comparison_PdPC} suggest that the relative tightness of the two types of bounds depends on the model.

\subsubsection{Multiple chemical cycles $(\delta = 0)$}\label{sec:deficiency_zero_numerical_multiple_cycles}

\begin{figure}[t]
\centering
\includegraphics[width=\columnwidth]{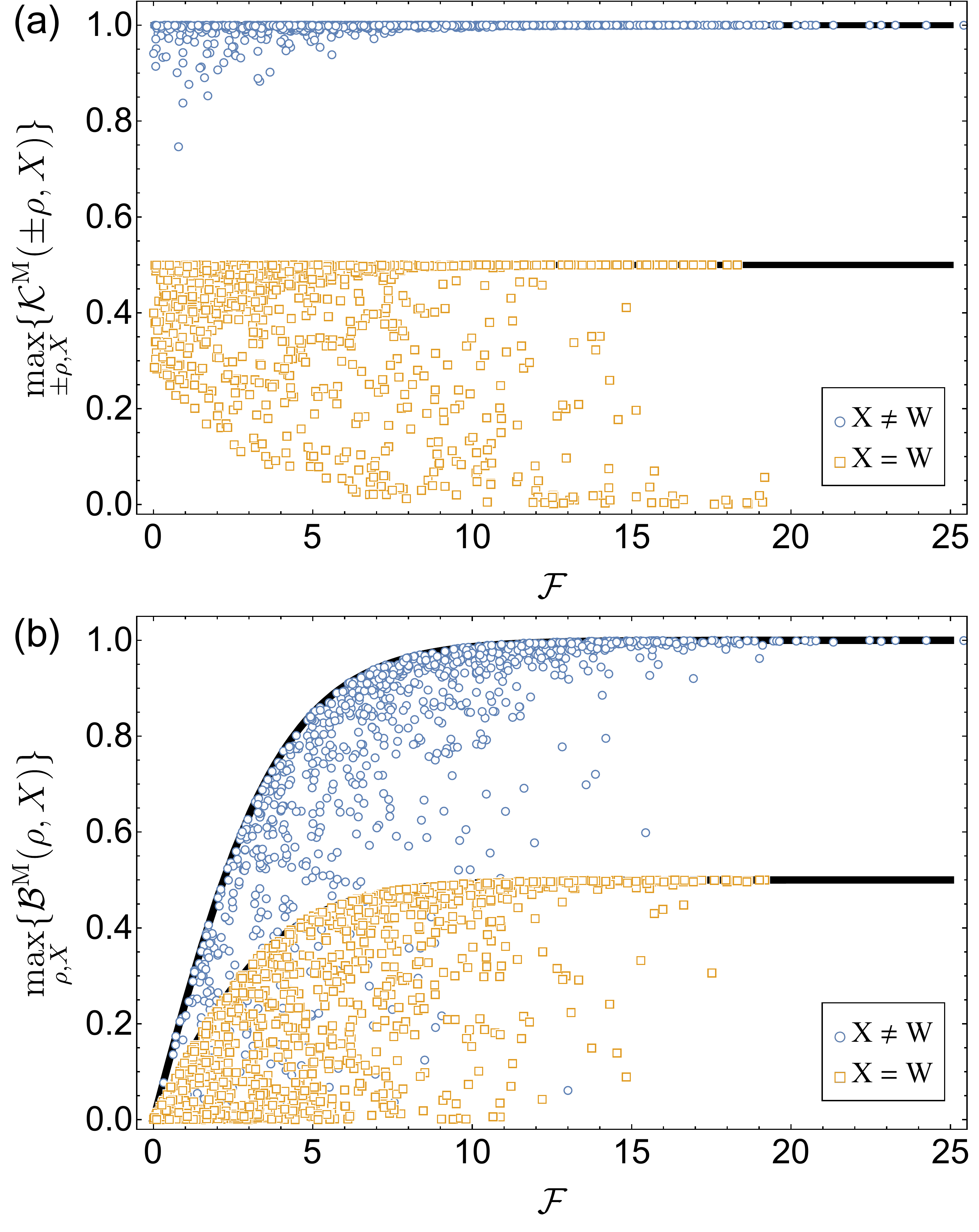}
\caption{The maximum quality factors among different choices of $\bm{(}\rho, X (\neq W)\bm{)}$ for the model in \eqref{eq:modified_PdPC2} at a fixed $(S_T, \{ k_{\pm \rho}\}_{\rho=1}^7)$ (circles).
The maximum quality factors for $X = W$ are calculated separately (squares).
Each symbol is obtained from a set of randomly sampled rate constants $\{ k_{\pm \rho}\}_{\rho=1}^7$ and the total concentration of substrates $S_T$, the logs of which are uniformly sampled within the range $[-5,5]$.
The number of each symbol in each panel is $10^3$.
The thick lines represent $1/2$ and $1$ in (a), and $(1/2)\tanh(\mathcal{F}/4)$ and $\tanh(\mathcal{F}/4)$ in (b).
}
\label{fig:modified_PdPC2}
\end{figure}

We further alter the modification-demodification cycle to include the effect of multiple chemical cycles: 
\begin{equation}\label{eq:modified_PdPC2}
\schemestart
    \chemfig{ES_2} \arrow{<=>[\tiny $k_{-3}$][\tiny $k_{+3}$]} \chemfig{E+P}
    \arrow{<=>[\tiny $k_{-2}$][\tiny $k_{+2}$]}[90] \chemfig{ES_1}
    \arrow{<=>[\tiny $k_{+1}$][\tiny $k_{-1}$]}[180] \chemfig{E + S}
    \arrow{<=>[\tiny $k_{-4}$][\tiny $k_{+4}$]}[270]
\schemestop\hspace{1cm}
\schemestart
    \chemfig{E} \arrow{<=>[\tiny $k_{+6}$][\tiny $k_{-6}$]} \chemfig{2W}
    \arrow{<=>[\tiny $k_{-7}$][\tiny $k_{+7}$]}[120] \chemfig{\emptyset}
    \arrow{<=>[\tiny $k_{-5}$][\tiny $k_{+5}$]}[240]
\schemestop.
\end{equation}
The enzyme $E$ can now also be reversibly converted into a waste form $W$, and both the enzyme and the waste species are in contact with chemical reservoirs.
Two chemical cycles are visible in the complex-reaction graph.
We denote the associated thermodynamic forces in the left and right linkage classes as $F_l$ and $F_r$.
The maximum thermodynamic force $\mathcal{F}$ \eqref{eq:maxF} is equal to $|F_l|$ ($|F_r|$) if the perturbed reaction is in the left (right, resp.) linkage class.
The total number of enzymes is no longer conserved, while the conservation law for substrates, $\Lambda_S = n_S + n_P + n_{ES_1} + n_{ES_2}$, remains.
Since the numbers of $E$ and $W$ are not limited, the stoichiometric compatibility class is infinitely large even if $\Lambda_S$ is small.
At the same time, the zero deficiency of the CRN ensures that the steady-state (positive) solution of the Chemical Rate Equations is unique~\cite{feinberg2019foundations}.
Instead of solving the Chemical Master Equation with small $\Lambda_S$, we examine the validity of the macroscopic version of trade-offs by considering quality factors ${\mathcal K}^{\rm M}(\pm\rho,X)$ and ${\mathcal B}^{\rm M}(\rho,X)$.
For the macroscopic concentrations, we denote the conserved total concentration of substrates by $S_T = [S]+[P]+[ES_1]+[ES_2]$.

We numerically calculated the maximum values of ${\mathcal K}^{\rm M}(\pm\rho, X)$ and ${\mathcal B}^{\rm M}(\rho,X)$ for different pairs $(\rho, X)$ for randomly-sampled combinations of system parameters $( S_T, \{ k_{\pm \rho}\}_{\rho=1}^7)$.
The numerical results are presented in Fig.~\ref{fig:modified_PdPC2} in which the cases $X = W$ and $X \neq W$ are treated separately.
The maximum quality factor of the wasted form of enzyme $W$ is bounded from above by 1/2 while those of other chemical species are bounded by 1.
Figure~\ref{fig:modified_PdPC2} and TABLE~\ref{tbl:modified_PdPC2} suggest the model in \eqref{eq:modified_PdPC2} obeys the macroscopic version of trade-offs ${\mathcal K}^{\rm M} \leq \alpha_1$ and ${\mathcal B}^{\rm M} \leq \alpha_2 \tanh(\mathcal{F}/4)$ with $\alpha_1 = \alpha_2 = 1/2$ for $X = W$ and $\alpha_1 = \alpha_2 = 1$ for $X\neq W$.

The suppressed response of $W$ is a nonlinear effect.
Since the steady-state distribution is a product form but no conservation law constrains $n_W$, the fluctuations of $n_W$ are decoupled from other chemical species and thus ${\rm Cov}_\Gamma\{n_W,n_j\} = {\rm Var}_\Gamma\{n_W\} \delta_{Wj}$.
As a result, the fluctuation-response relation \eqref{eq:FRR} for $W$ simplifies to 
\begin{equation}
    \frac{\partial \langle n_W \rangle_\Gamma}{\partial \lambda}
    = \frac{1}{2} {\rm Var}_\Gamma \{n_W \}  \frac{\partial}{\partial \lambda} \ln [W].
\end{equation}
which in turn leads to trade-offs with $\alpha_1 = \alpha_2 = 1/2$. Since a compatibility class is composed of a single connected component in this model, applying the matrix tree theorem, we can actually prove trade-offs for $E$ and $W$, whose numbers are not conserved for any $\Lambda_S$.

The above statements can be generalized.
For a deficiency zero CRN, if the number of a chemical species $A$ is not conserved and a complex $z A$ with a positive integer $z$ is included in the CRN, the fluctuation-response relation \eqref{eq:FRR} for $A$ can be written as $\partial_\lambda \langle n_A \rangle_\Gamma = (1/z){\rm Var}_\Gamma\{n_A\} \partial_\lambda \ln [A]^z$.
By noticing $\Psi_{zA}([\bm{X}]_{\rm ss}) = [A]^z$ and $\Psi_\emptyset([\bm{X}]_{\rm ss}) = 1$ and applying the matrix-tree theorem, we can derive trade-offs $|k_{\pm\rho}\partial \langle n_A \rangle_\Gamma/\partial k_{\pm\rho}| \leq {\rm Var}_\Gamma\{n_A\}/z$ and $|\partial \langle n_A \rangle_\Gamma/\partial B_{\rho}| \leq ({\rm Var}_\Gamma\{n_A\}/z) \tanh(\mathcal{F}/4)$, provided the complexes $\emptyset$ and $zA$ are in the same linkage class.

\begin{table}[t]
\centering
\begin{tabular}{cccccccc}
\toprule
\multicolumn{1}{c}{} & \multicolumn{7}{c}{\textbf{chemical reaction} $(\rho)$} \\
\cmidrule(rl){2-8} 
$X$ &  $\pm 1$ & ~ $\pm 2$ & ~ $\pm 3$ & ~$\pm 4$ & ~$\pm 5$ & ~$\pm 6$ & ~$\pm 7$ \\
\hline\hline
$E$ & (0,~0) & ~ (0,~0) & ~ (0,~0) & ~ (0,~0) & ~ (1,~1) & ~ (1,~1) & ~ (1,~1) \\
$S$ & (1,~1) & ~ (1,~1) & ~ (1,~1) & ~ (1,~1) & ~ (1,~1) & ~ (1,~1) & ~ (1,~1) \\
$P$ & (1,~1) & ~ (1,~1) & ~ (1,~1) & ~ (1,~1) & ~ (1,~1) & ~ (1,~1) & ~ (1,~1) \\
$ES_1$ & (1,~1) & ~ (1,~1) & ~ (1,~1) & ~ (1,~1) & ~ (1,~1) & ~ (1,~1) & ~ (1,~1) \\
$ES_2$ & (1,~1) & ~ (1,~1) & ~ (1,~1) & ~ (1,~1) & ~ (1,~1) & ~ (1,~1) & ~ (1,~1) \\
$W$ & (0,~0) & ~ (0,~0) & ~ (0,~0) & ~ (0,~0) & ~ ($\frac{1}{2}$,~$\frac{1}{2}$) & ~ ($\frac{1}{2}$,~$\frac{1}{2}$) & ~ ($\frac{1}{2}$,~$\frac{1}{2}$) \\
\bottomrule
\end{tabular}
\caption{Numerically estimated values of ($\alpha_1$, $\alpha_2$) to each given pair $(\rho,X)$ for the model in \eqref{eq:modified_PdPC2}.}
\label{tbl:modified_PdPC2}
\end{table}

\section{Deficiency Nonzero Chemical Reaction Networks}\label{sec:deficiency_nonzero}

Analytical approaches are further limited for deficiency nonzero CRNs since the general form of the steady-state distribution is unknown.
So, we exclusively rely on numerical methods here.
The validity of trade-offs \eqref{eq:main1} and \eqref{eq:main2} are examined by calculating the quality factors defined in the previous section for two models with different features.

\subsection{Multiple chemical cycles $(\delta \neq 0)$}

The Sch\"ogl model illustrated in Sec.~\ref{sec:illustration} is a deficiency nonzero model with a single chemical cycle, for which the trade-offs between fluctuation and response are proven.
In order to see the effect of multiple thermodynamic forces, we consider a deficiency nonzero model with two independent chemical cycles, whose complex-reaction graph is given by 
\begin{equation}\label{eq:model_deficiency_nonzero_two_cycles}
\schemestart
    \chemfig{X} \arrow{<=>[\tiny $k_{+2}$][\tiny $k_{-2}$]} \chemfig{2X}
    \arrow{<=>[\tiny $k_{-3}$][\tiny $k_{+3}$]}[120] \chemfig{\emptyset}
    \arrow{<=>[\tiny $k_{-1}$][\tiny $k_{+1}$]}[240]
\schemestop.
\end{equation}
Apart from the cycle visible in the complex-reaction graph with thermodynamic force $F_1 = \ln\bm{(}(k_{+1}k_{+2}k_{+3})/(k_{-1}k_{-2}k_{-3})\bm{)}$, an independent cycle with thermodynamic force $F_2 = \ln\bm{(}(k_{-1}k_{+2})/(k_{+1}k_{-2})\bm{)}$ is further identified by analysis of the stoichiometric matrix.
Along with $F_1$ and $F_2$, two more cycles characterized by thermodynamics forces $F_1 - F_2 = \ln\bm{(}(k_{+1}^2 k_{+3})/(k_{-1}^2 k_{-3})\bm{)}$ and $F_1 + F_2 = \ln\bm{(}(k_{+2}^2k_{+3})/(k_{-2}^2k_{-3})\bm{)}$ constitute the entire set of cycles that reside in the microscopic state space.
The maximum thermodynamic force $\mathcal{F}$ depends on the perturbed edge \eqref{eq:maxF}.
For example, if the chemical reaction $\emptyset \rightleftharpoons X$ is perturbed, $\mathcal{F} = \max\{ |F_1|, |F_2|, |F_1-F_2| \}$.

\begin{figure}[t]
\centering
\includegraphics[width=\columnwidth]{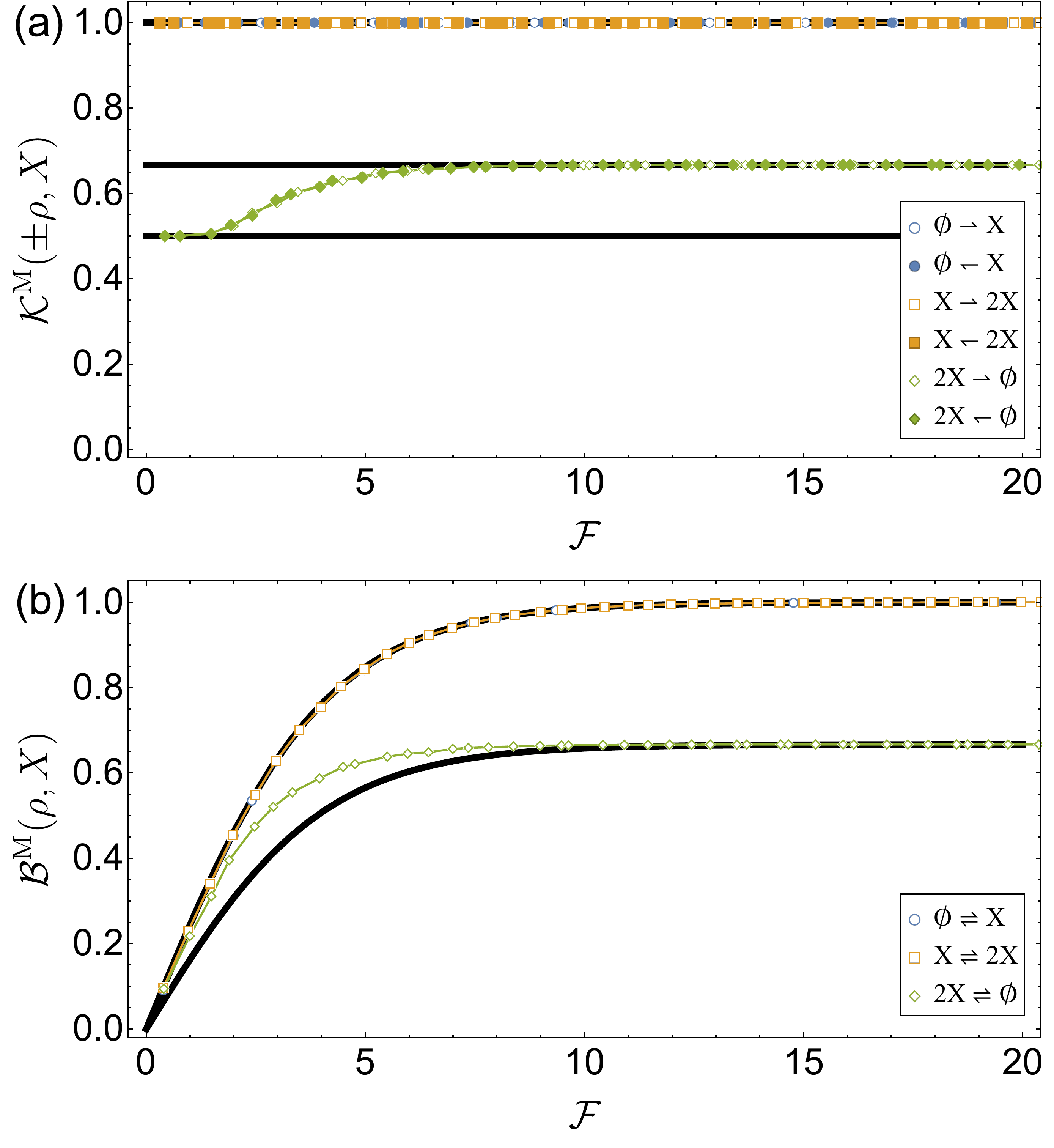}
\caption{Quality factors for the model in \eqref{eq:model_deficiency_nonzero_two_cycles}.
Raw data are calculated from $10^6$ sets of randomly sampled rate constants, the logs of which are uniformly sampled over the range $[-10,10]$.
The symbols represent the maximum values of quality factors selected from the sampled data within range $\mathcal{F} \in [x,x+0.5]$ where $x \in \{ 0, 0.5, 1, \cdots, 19.5 \} $.
The thick lines represent $\{ 1/2, 2/3, 1\}$ in (a) and $\{ (2/3)\tanh(\mathcal{F}/4), \tanh(\mathcal{F}/4) \}$ in (b).}
\label{fig:deficiency_nonzero_CRN_two_cycles}
\end{figure}

The number of chemical species $X$ is not conserved with no upper bound.
Thus the only stoichiometric compatibility class is infinitely large.
We examine the validity of the macroscopic version of the trade-offs \eqref{eq:main1} and \eqref{eq:main2}.
The steady-state concentration $[X]_{\rm ss}$ is the unique positive solution of the quadratic equation
\begin{equation}
    (k_{+1} + 2k_{-3}) + (-k_{-1} + k_{+2})x - (k_{-2} + 2k_{+3}) x^2 = 0.
\end{equation}
We numerically calculate the quality factors ${\mathcal K}^{\rm M}(\pm\rho, X)$ and $\mathcal{B}^{\rm M}(\rho,X)$ for a fixed set of rate constants $\{ k_{\pm \rho}\}_{\rho=1}^3$.
The logs of the rate constants are randomly sampled from a uniform distribution over the range $[-10,10]$.
From these $10^6$ data points, the maximum response is singled out from each interval $\mathcal{F} \in [x,x+0.5]$ with $x \in \{ 0, 0.5, 1, \cdots, 19.5 \}$ and plotted in Fig.~\ref{fig:deficiency_nonzero_CRN_two_cycles}.
Numerical results are consistent with the macroscopic version of the trade-offs.
The prefactors and the function $g(\mathcal{F})$ vary for different choices of the perturbed chemical reaction. 
For example, when the chemical reaction $2X \rightleftharpoons \emptyset$ is perturbed, the quality factors do not exceed $2/3$ and $g(\mathcal{F})$ differs from $\tanh(\mathcal{F}/4)$.
Unlike in the deficiency zero CRNs without a conservation law (see Sec.~\ref{sec:deficiency_zero_numerical_multiple_cycles}), the effect of nonlinearity is not to simply reduce the response function by an integer factor.

\subsection{Multiple conservation laws $(\delta \neq 0)$}\label{sec:GB}

Our inspiration here is the Goldbeter-Koshland model for the PdPC.
In its original formulation, a substrate can exist in an unmodified $S$ form and a modified or phosphorylated form $P$.
Phosphorylation is catalyzed by a kinase enzyme $E_1$, whereas a phosphatase enzyme $E_2$ catalyzes dephosphorylation.
By introducing two intermediate species $E_1S$ and $E_2P$, Goldbeter and Koshland considered two chains of irreversible chemical reactions~\cite{goldbeter1981amplified}
\begin{equation}\label{eq:PdPC_model}
\begin{aligned}
    E_1 + S \rightleftharpoons E_1 S \rightharpoonup E_1 + P, \\
    E_2 + P \rightleftharpoons E_2 P \rightharpoonup E_2 + S
\end{aligned}
\end{equation}
as a mechanism to explain switch-like behavior observed in some biological scenarios.
The high sensitivity, called zero-order ultrasensitivity~\cite{ferrell2014ultrasensitivity}, is observed in a regime where the total numbers of both enzymes are much lower than that of the substrates and the chemical cycle is highly irreversible~\cite{qian2003thermodynamic}.
We will examine the emergence of this high sensitivity from the point of view of the trade-offs.

The irreversible chemical reactions of the model imply infinite entropy production and are thermodynamically implausible.
In line with \cite{qian2003thermodynamic}, we consider a reversible version with complex-reaction graph
\begin{equation}\label{eq:reversible_PdPC_model}
\begin{aligned}
    E_1 + S \xrightleftharpoons[k_{-1}]{k_{+1}}
    E_1 S \xrightleftharpoons[k_{-2}]{k_{+2}}
    E_1 + P, \\
    E_2 + P \xrightleftharpoons[k_{-3}]{k_{+3}}
    E_2 P \xrightleftharpoons[k_{-4}]{k_{+4}}
    E_2 + S.
\end{aligned}
\end{equation}
The deficiency of the model is one.
The stoichiometric compatibility class is specified by three conservation laws:
the total numbers of the respective enzymes, $\Lambda_{E_1} = n_{E_1} + n_{E_1S}$ and $\Lambda_{E_2} = n_{E_2} + n_{E_2P}$, as well as the total number of substrates, $\Lambda_S = n_{S} + n_{P} + n_{E_1S} + n_{E_2P}$.
The thermodynamic force associated with the only chemical cycle invisible in the complex-reaction graph is $F = \ln\bm{(}\prod_{i=1}^4 (k_{+i}/k_{-i})\bm{)}$, which is the chemical potential difference between ATP and ADP+P\textsubscript{i} in the unit of $k_B T$.

\begin{figure}[t]
\centering
\includegraphics[width=\columnwidth]{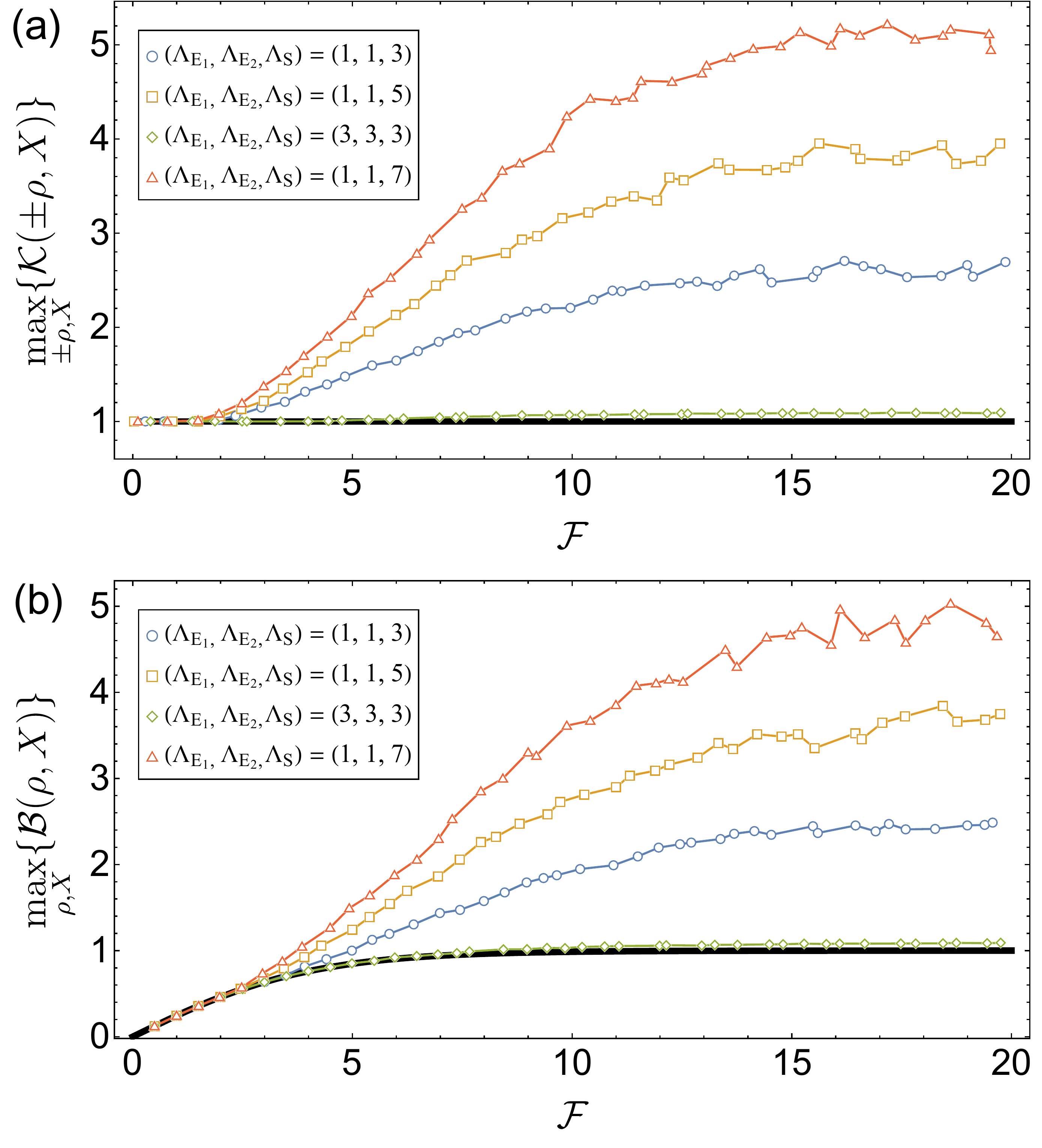}
\caption{The maximum quality factors among different choices of ($\rho$, $X$) for the model in \eqref{eq:reversible_PdPC_model} with $(\Lambda_{E_1} , \Lambda_{E_2}, \Lambda_S) \in \{ (1,1,3), ~(1,1,5),~ (3,3,3), ~ (1,1,7) \}$ at a fixed $(\{ k_{\pm \rho}\}_{\rho=1}^4, \Omega)$.
Raw data are calculated from $10^6$ sets of randomly sampled rate constants $\{ k_{\pm \rho}\}_{\rho=1}^4$ and system size $\Omega$, the logs of which are uniformly sampled within the range $[-5,5]$.
The symbols represent the maximum values of quality factors selected out of raw data within range $\mathcal{F} \in [x,x+0.5]$ where $x \in \{ 0, 0.5, 1, \cdots, 19.5 \} $
The thick lines represent $1$ in (a), and $\tanh(\mathcal{F}/4)$ in (b).}
\label{fig:PdPC}
\end{figure}

In order to examine the validity of the trade-offs \eqref{eq:main1} and \eqref{eq:main2}, we numerically calculate the quality factors $\mathcal{K}(\pm \rho, X)$ and $\mathcal{B}(\rho, X)$ at $10^6$ fixed values of $(\{ k_{\pm \rho}\}_{\rho=1}^4, \Omega )$.
The logs of the rate constants $\{ k_{\pm \rho}\}_{\rho=1}^4$ and system volume $\Omega$ are randomly sampled from a uniform distribution over the range $[-5,5]$, from which each $\mathcal{K}(\pm \rho, X)$ and $\mathcal{B}(\rho, X)$ are determined for all pairs $(\rho, X)$.
For each value of the system parameters, we then select the maximum of $\mathcal{K}(\pm \rho, X)$ and $\mathcal{B}(\rho, X)$ over all pairs $(\rho, X)$ to obtain one sample of the max-quality factors.
Plotted in Figure~\ref{fig:PdPC} is then the maximum quality factor within each interval $\mathcal{F} \in [x,x+0.5]$ for $x \in \{ 0, 0.5, 1, \cdots, 19.5 \}$.
The response depends on the total numbers of enzymes and substrates, $(\Lambda_{E_1}, \Lambda_{E_2}, \Lambda_S) \in \{ (1,1,3), ~(1,1,5),~ (3,3,3), ~ (1,1,7) \}$, which also determine the respective size of the stoichiometric compatibility classes, $\{ 12, 20, 20, 28 \}$.
Two features distinct from deficiency zero CRNs can be observed in Fig.~\ref{fig:PdPC}.

We first compare the cases $(\Lambda_{E_1} , \Lambda_{E_2}, \Lambda_S) = (1,1,5)$ and $(3,3,3)$ (orange and green), which have the same numbers of microscopic states in the stoichiometric compatibility class, 20.
Although the numbers of microscopic states are the same, the structural differences in the microscopic state space lead to a different quantitative relationship between response and fluctuations.
While the maximum values of the quality factors are slightly greater than 1 when the number of enzymes is comparable with that of the substrates, they appear to grow to a much larger value with thermodynamic force when the enzymes are scarce.
This high response in Fig.~\ref{fig:PdPC} observed when the enzymes are scarce is reminiscent of zero-order ultrasensitivity; however, in the biochemical literature ultrasensitivity is defined by the response to a change in the total numbers of enzymes $\Lambda_{E_1}$ and/or $\Lambda_{E_2}$~\cite{ferrell2014ultrasensitivity}, which is different from what we consider here.
With that caveat in mind, a recent theoretical study has shown that in the limit that substrate concentration goes to infinity (infinitely scarce enzymes), the log-sensitivity to changes in the total enzyme concentration can actually grow as fast as exponential with thermodynamic driving~\cite{owen2022ultra}.
This prediction appears consistent with our observation that our quality factors grow over a wider range of thermodynamic forces when there are less enzymes.

Next, we compare the cases $(\Lambda_{E_1} , \Lambda_{E_2}, \Lambda_S) = (1,1,3), ~ (1,1,5)$, and $(1,1,7)$.
On the one hand, the quality factors appear to obey the trade-offs \eqref{eq:main1} and \eqref{eq:main2} for a fixed $\Lambda_S$.
On the other hand, as the total number of substrates increases, the maximum values of the quality factors increase as well.
This suggests that the prefactors $\alpha_{1,2}$ in the trade-offs depend on system details, such as the ratio between the numbers of substrates and enzymes, which deform the structure of the stoichiometric compatibility class.

\section{Concluding Remarks}\label{sec:conclusion}
In this work, we investigated fluctuation-response trade-offs for CRNs in nonequilibrium steady-states.
We proved the trade-offs, \eqref{eq:main1} and \eqref{eq:main2}, for linear CRNs (Sec.~\ref{sec:deficiency_zero}) and a class of nonlinear CRNs with a single chemical species (Sec.~\ref{sec:illustration}).
For deficiency zero nonlinear CRNs, the response is still bounded by fluctuations and thermodynamic forces via the inequalities \eqref{eq:deficiency_zero_ineq1} and \eqref{eq:deficiency_zero_ineq2}, though we have not been able to connect them to the compact trade-offs \eqref{eq:main1} and \eqref{eq:main2}.
Nevertheless, numerical calculations for nonlinear models suggest that the trade-offs \eqref{eq:main1} and \eqref{eq:main2} may continue to hold.
To support this observation, we have analyzed various models with differing aspects, such as whether there are conservation laws, the number of independent thermodynamic forces, the number of chemical species, and the deficiency of the CRN.
Regardless of different aspects, deficiency zero CRNs appear to share the following properties: (i) $\alpha_1 = \alpha_2$, (ii) $g(\mathcal{F}) = \tanh(\mathcal{F}/4)$, and (iii) the prefactors do not depend on the size of the stoichiometric compatibility class.
In contrast, numerical results for deficiency-nonzero CRNs show that  the prefactors depend on system details that affect the structure of the stoichiometric compatibility class (Sec.~\ref{sec:deficiency_nonzero}).

Trade-offs between number fluctuations and response suggest limits to the accuracy and sensitivity of biochemical systems.
Thus, our results pave the way for identifying fundamental limitations to nonequilibrium biochemical processes.
Moving forward new analytic techniques may be required, especially for nonlinear CRNs where the general form of the steady-state distribution is unknown. 
Such advances would allow us to discern how characteristics of CRNs, such as conservation laws or deficiency, restrict the topological structure of the microscopic state space and thus affect the response.

\section*{Acknowledgements}
H.-M.C. was supported in part by a KIAS Individual Grant (PG089401) at Korea Institute for Advanced Study.
J.M.H. acknowledges support from the National Science Foundation under Grant No.~2142466.

\appendix

\section{Numerical Methods}\label{sec:diffusion_approximation}

When the size of the isolated connected component of a stoichiometric compatibility class is sufficiently small, we first numerically solve the Chemical Master Equation \eqref{eq:CME1} for the steady-state distribution,  $\hat{\mathcal{L}}\pi_\Gamma(\bm{n}) = 0$ with normalization condition $\sum_{\bm{n}\in\Gamma} \pi_\Gamma({\bm n}) = 1$, which is a system of linear equations.
To determine the steady-state response  $\partial_\lambda \pi_\Gamma({\bm n})$, we differentiate the Chemical Master Equation with respect to the perturbation parameter $\lambda \in \{ \ln k_{+\rho}, \ln k_{-\rho}, B_\rho \}$, which leads to a system of inhomogeneous linear equations for the response 
\begin{equation}\label{eq:linear_eq_response}
    \hat{\mathcal{L}} \frac{\partial \pi_\Gamma(\bm{n})}{\partial \lambda} + \frac{\partial\hat{\mathcal{L}}}{\partial \lambda} \pi_\Gamma(\bm{n}) = 0,
\end{equation}
which are solved numerically together with the condition $\sum_{\bm{n}\in\Gamma} \partial_\lambda \pi_\Gamma({\bm n})=0$ that is required by normalization of the steady-state distribution.

When the system size is large, numerical solutions to the Chemical Master Equation \eqref{eq:CME1} are infeasible.  
In these scenarios, we first numerically solve the Chemical Rate Equation  for the steady-state concentration $[{\bm X}_{\rm ss}]$.
To determine the steady-state response $\partial_\lambda [{\bm X}_{\rm ss}]$, we develop a nonlinear equation for it by differentiating the Chemical Rate Equation
\begin{equation}\label{eq:nonlinear_eq_response}
\begin{aligned}
    &\sum_{\rho \in \mathcal{R}}
    \Delta\bm{\nu}_\rho
    \left( \frac{\partial k_{+\rho}}{\partial \lambda} [\bm{X}]_{\rm ss}^{\bm{\nu}^+_\rho} 
    - \frac{\partial k_{-\rho}}{\partial \lambda} [\bm{X}]_{\rm ss}^{\bm{\nu}^-_\rho} \right) \\
    & + \sum_{\rho \in \mathcal{R}}
    \Delta\bm{\nu}_\rho
    \left( k_{+\rho} \frac{\partial [\bm{X}]_{\rm ss}^{\bm{\nu}^+_\rho}}{\partial \lambda}  
    - k_{-\rho} \frac{\partial [\bm{X}]_{\rm ss}^{\bm{\nu}^-_\rho}}{\partial \lambda}  \right) = \bm{0},
\end{aligned}
\end{equation}
which allows us to numerically determine the response.
However, the Chemical Rate Equation does not encode any information about the fluctuations.

To ascertain the scaled variance $D_{i}$, we use Van Kampen's system size expansion, also known as the Linear Noise Approximation~\cite{van2007stochastic}, to approximate the Chemical Master Equation as a diffusion process, which we review here.
In this approximation, the numbers of chemical species are assumed to be split as $\bm{n} = \Omega[\bm{X}] + \Omega^{1/2}\delta\bm{n}$ in terms of the deterministic solution $[\bm{X}]$ of the Chemical Rate Equation \eqref{eq:CRE1} and Gaussian fluctuations $\delta\bm{n}$.
Such a decomposition of $\bm{n}$ allows us to approximate the mean scaled chemical reaction rates as
\begin{equation}
\begin{aligned}
    \frac{\langle w_{\pm\rho}(\bm{n}) \rangle}{\Omega}
    &= \frac{\langle w_{\pm\rho}(\Omega[\bm{X}] + \Omega^{1/2} \delta\bm{n}) \rangle}{\Omega} \\
    & \approx \frac{w_{\pm\rho}(\Omega[\bm{X}])}{\Omega}
    + \left. \frac{\langle \delta \bm{n} \rangle \cdot \nabla_{\bm{n}} w_{\pm\rho}(\bm{n})}{\Omega^{1/2}} \right|_{\bm{n}=\Omega[\bm{X}]} \\
    &\approx k_{\pm\rho}[\bm{X}]^{\bm{\nu}_\rho^\pm}
    + \Omega^{-1/2} \langle \delta\bm{n} \rangle \cdot
    \bm{\nabla}_{[\bm{X}]} k_{\pm\rho} [\bm{X}]^{\bm{\nu}^\pm_\rho}.
\end{aligned}
\end{equation}
As a result, we can approximate the time-evolution of the mean and covariance of $\delta \bm{n}$ as
\begin{equation}
\begin{aligned}
    \frac{d\langle \delta\bm{n}_t \rangle}{dt}
    &= \Omega^{1/2} \left( \frac{1}{\Omega} \frac{d\langle \bm{n}_t \rangle}{dt} - \frac{d[\bm{X}]_t}{dt} \right) \\
    & = \Omega^{1/2} \sum_{\rho\in\mathcal{R}} \Delta\bm{\nu}_\rho \left( \frac{ \langle w_{+\rho}(\bm{n}_t) \rangle - \langle w_{-\rho}(\bm{n}_t) \rangle}{\Omega} \right) \\
    & ~~~ - \Omega^{1/2} \sum_{\rho\in\mathcal{R}} \Delta\bm{\nu}_\rho \left( k_{+\rho} [\bm{X}]^{\bm{\nu}_\rho^+} - k_{-\rho} [\bm{X}]^{\bm{\nu}_\rho^-}  \right) \\
    & \approx \langle \delta\bm{n}_t \rangle \cdot \bm{\nabla}_{[\bm{X}_t]} \sum_{\rho\in\mathcal{R}} \Delta\bm{\nu}_\rho \left( k_{+\rho} [\bm{X}]_t^{\bm{\nu}^+_\rho} - k_{-\rho} [\bm{X}]_t^{\bm{\nu}^-_\rho} \right) \\
    & \equiv \mathsf{L} \cdot \langle \delta\bm{n}_t \rangle
\end{aligned}
\end{equation}
and
\begin{equation}
\begin{aligned}
    & \frac{d {\rm Cov}_t\{ \delta n_i, \delta n_j \}}{dt}
    = \frac{1}{\Omega} \frac{d{\rm Cov}_t\{ n_i, n_j \}}{dt} \\
    & = \frac{1}{\Omega} \sum_{\rho \in \mathcal{R}} S_{i\rho} S_{j\rho} \left\{ w_{+\rho}(\bm{n}_t) + w_{-\rho}(\bm{n}_t) \right\} \\
    & \approx \sum_{\rho \in \mathcal{R}} S_{i\rho} S_{j\rho} \left( k_{+\rho}[\bm{X}]_t^{\bm{\nu}^+_\rho} + k_{-\rho}[\bm{X}]_t^{\bm{\nu}^-_\rho} \right)
    \equiv N_{ij}.
\end{aligned}
\end{equation}
The last equalities define the relaxation matrix $\mathsf{L}$ and noise matrix $\mathsf{N}$.
Together these equations imply that the Gaussian dynamics of $\delta\bm{n}$ can be captured with the linear Langevin equation
\begin{equation}
    \frac{d(\delta\bm{n}_t)}{dt}
    = \mathsf{L} \cdot \delta\bm{n}_t + \bm{\zeta}_t,
\end{equation}
where $\bm{\zeta}_t$ is a multivariate zero-mean Gaussian white noise with covariance $\mathsf{N}$.
As a result, the steady-state covariance $V_{ij} = {\rm Cov}\{n_i, n_j\}$ of a linear Langevin equation can be obtained as the solution of the algebraic equation~\cite{gardiner2009stochastic}
\begin{equation}\label{eq:matrix_eq}
    \mathsf{L} \cdot \mathsf{V} + \mathsf{V} \cdot \mathsf{L}^{\rm T} + \mathsf{N} = 0,
\end{equation}
which we solve numerically.  
The scaled variance is then obtained as $D_{i} = {\rm Var}\{ n_i \}/\Omega = {\rm Var}\{ \delta n_i \}$.

\section{Trade-offs for General Deficiency Zero CRNs}\label{sec:deficiency_zero_nonlinear_CRNs}

The structure of the equation $\mathsf{A}\cdot\bm{\Psi}([\bm{X}]_{\rm ss}) = \bm{0}$ that determines $[\bm{X}]_{\rm ss}$ allows us to find restrictions on the response functions for the deterministic concentrations.
The vector $\bm{\Psi}([\bm{X}]_{\rm ss})$ lies in the right null space of the Laplacian matrix $\mathsf{A}$.
One useful basis of the right null space of the Laplacian matrix is provided by the matrix-tree theorem (MTT)~\cite{hill1966studies,gunawardena2012linear,owen2020universal}.

To exploit this connection, a few graph-theoretical notions are needed.
A complex-reaction graph may consist of multiple linkage classes, which we denote as $g_p$ with $p \in \{ 1, 2, \cdots, \ell \}$.
The MTT is then built out of rooted spanning trees of each $g_p$, which are connected subgraphs that contain no cycles, include all vertices, and have each edge oriented towards a specific vertex, $\bm{y}_l$.
A weight is then assigned to each tree by noting that each directed edge is endowed with a rate constant from the corresponding chemical reaction: the weight of the subgraph is then product of the associated rate constants.
We denote the sum of the weights of every spanning tree rooted at vertex $\bm{y}_l$ by $T_l$.
The MTT then states that the vectors $\bm{T}^{(p)}$ with components
\begin{equation}
    T^{(p)}_l = \begin{cases}
        T_l & {\rm if} ~ \bm{y}_l \in g_p, \\
        0 & {\rm otherwise},
    \end{cases}
\end{equation}
lie in the right null space of the Laplacian matrix, i.e., $\mathsf{A} \cdot \bm{T}^{(p)} = \bm{0}$ for all $p \in \{ 1,2,\cdots,\ell \}$~\cite{hill1966studies,gunawardena2012linear,owen2020universal}.
Since the $\bm{T}^{(p)}$ with different $p$ are linearly independent of each other, the collection of $\bm{T}^{(p)}$ for $p \in \{ 1, 2, \cdots, \ell \}$ can serve as a basis of the right null space of $\mathsf{A}$.
It follows that $\Psi_l([\bm{X}]_{\rm ss}) / \Psi_m([\bm{X}]_{\rm ss}) = T_l / T_m$ if $\bm{y}_l$ and $\bm{y}_m$ are in the same linkage class.
The ratios for two vertices belonging to different linkage classes are determined by constraints imposed by conservation laws.

Now, due to the product form of the steady-state distribution \eqref{eq:product_form}, the static response of the log-ratio of steady-state probabilities to a perturbation is given by
\begin{equation}\label{eq:FRR_deficiency_zero_CRNs}
    \frac{\partial}{\partial \lambda}
    \ln \frac{\pi_\Gamma(\bm{n})}{\pi_\Gamma(\bm{n}')} 
    = \sum_{i\in\mathcal{S}} (n_i - n_i') \frac{\partial}{\partial \lambda} \ln [X_i]_{\rm ss}.
\end{equation}
In order to take advantage of the graph-theoretic tools, \eqref{eq:FRR_deficiency_zero_CRNs} needs to be rearranged to a form involving the $\Psi_i([\bm{X}]_{\rm ss})$, which can be replaced with spanning trees.
To this end, we change the basis from number, and instead identify each microscopic state by the number of (independent) reactions required to arrive at it.
We accomplish this by noting the matrix $\mathsf{S}'$, defined in \eqref{eq:reduced_stoichiometric_matrix}, converts the difference of state vectors as $\bm{n} - \bm{n}' = \mathsf{S}'\cdot \bm{\mu}(\bm{n},\bm{n}')$ into $\bm{\mu}(\bm{n},\bm{n}')$ vector whose components are the number of each reaction required to move from ${\bm n}'$ to ${\bm n}$.
Note that the decomposition of independent and dependent reactions used to construct $\mathsf{S}'$ is not unique.
Substituting this change of basis into \eqref{eq:FRR_deficiency_zero_CRNs}, coupled with the identity $\sum_{j\in\mathcal{S}} S_{j\rho}' \ln [X_j] = \ln [\bm{X}]_{\rm ss}^{\Delta \bm{\nu}_\rho} = \ln ([\bm{X}]_{\rm ss}^{\bm{\nu}_\rho^-}/[\bm{X}]_{\rm ss}^{\bm{\nu}_\rho^+})$, leads to
\begin{equation}\label{eq:FRR_deficiency_zero_CRNs2}
\begin{aligned}
    \frac{\partial}{\partial \lambda}
    \ln \frac{\pi_\Gamma(\bm{n})}{\pi_\Gamma(\bm{n}')}
    & = \sum_{\rho\in\mathcal{R}'} \mu_\rho(\bm{n},\bm{n}')
    \frac{\partial}{\partial \lambda}
    \ln \frac{[\bm{X}]_{\rm ss}^{\bm{\nu}_{\rho}^-}}
    {[\bm{X}]_{\rm ss}^{\bm{\nu}_{\rho}^+}} \\
    & = \sum_{\rho\in\mathcal{R}'} \mu_\rho(\bm{n},\bm{n}') \frac{\partial}{\partial \lambda} \ln \frac{\Psi_{{\bm \nu}_\rho^-}([{\bm X}_{\rm ss}])}
    {\Psi_{{\bm \nu}_\rho^+}([{\bm X}_{\rm ss}])} \\
    & = \sum_{\rho\in\mathcal{R}'} \mu_\rho(\bm{n},\bm{n}') \frac{\partial}{\partial \lambda} \ln \frac{T_{{\bm \nu}_\rho^-}}{T_{{\bm \nu}_\rho^+}}
\end{aligned}
\end{equation}
where sum runs over independent reactions, $\mathcal{R}' = \{1,2,\cdots,s\}$. Note the complexes $\bm{\nu}_\rho^-$ and $\bm{\nu}_\rho^+$ are always in the same linkage class since they are connected via the $\rho$-th chemical reaction.

As a result, the analysis reduces to bounding the sensitivity of ratios of spanning tree weights.
The bounds to the specific perturbations $\partial/\partial \lambda = k_{\pm\rho}\partial/\partial k_{\pm\rho}$ and $\partial/\partial \lambda = \partial/\partial B_{\rho}$, now follow directly from the methods developed in \cite{owen2020universal}, with the results
\begin{align}\label{eq:MTT_inequality1}
   & \left| k_{\pm\rho} \frac{\partial}{\partial k_{\pm\rho}}
   \ln \frac{T_{\bm{\nu}_{\sigma^-}}}
   {T_{\bm{\nu}_{\sigma^+}}} \right|
   \leq 1,\\
\label{eq:MTT_inequality2}
    & \left| \frac{\partial}{\partial B_\rho} 
    \ln \frac{T_{\bm{\nu}_{\sigma^-}}}
    {T_{\bm{\nu}_{\sigma^+}}} \right|
    \leq \tanh\left( \frac{\mathcal{F}}{4} \right).
\end{align}
Applying the inequalities \eqref{eq:MTT_inequality1} and \eqref{eq:MTT_inequality2} to \eqref{eq:FRR_deficiency_zero_CRNs2}, we obtain the bounds,
\begin{align}\label{eq:deficiency_zero_ratio_bound}
&    \left| k_{\pm\rho} \frac{\partial}{\partial k_{\pm\rho}}
    \ln \frac{\pi_\Gamma(\bm{n})}{\pi_\Gamma(\bm{n}')}  \right|
    \leq \sum_{\rho\in\mathcal{R}'}
    \left| \mu_\rho(\bm{n},\bm{n}') \right|,\\
   & \left| \frac{\partial}{\partial B_{\rho}}
    \ln \frac{\pi_\Gamma(\bm{n})}{\pi_\Gamma(\bm{n}')} \right|
    \leq \sum_{\rho\in\mathcal{R}'}
    \left| \mu_\rho(\bm{n},\bm{n}') \right|
    \tanh\left(\frac{\mathcal{F}}{4}\right).
\end{align}
The bound \eqref{eq:deficiency_zero_ratio_bound} is reminiscent of \eqref{eq:birth-death_ratio_bound} since $\mu_\rho(\bm{n},\bm{n}')$ is a measure of the distance between two states $\bm{n}$ and $\bm{n}'$.
If the dynamics are a birth-death process, then the chemical reaction changes the number of chemical species by one, and the bound \eqref{eq:deficiency_zero_ratio_bound} coincides with \eqref{eq:birth-death_ratio_bound}.
Nevertheless, the derivations are independent.
The class of CRNs considered in Sec.~\ref{sec:illustration} may have a nonzero deficiency (e.g.\ the Schl\"ogl model has a deficiency of one), whereas \eqref{eq:deficiency_zero_ratio_bound} applies only to deficiency zero CRNs.

Lastly, we derive trade-offs between the response of the mean number of a chemical species and fluctuations.
The response of the conditional mean number $\langle n_i \rangle_\Gamma = \sum_{\bm{n} \in\Gamma} n_i  \pi_\Gamma(\bm{n})$ on an isolated connected component $\Gamma$ is given by multiplying $n_i \pi_\Gamma(\bm{n}) \pi_\Gamma(\bm{n}')$ by both sides of \eqref{eq:FRR_deficiency_zero_CRNs2} and summing over $\bm{n}, \bm{n}' \in \Gamma$.
Using the identity $\sum_{\bm{n}'} \pi_\Gamma(\bm{n}') \partial \ln\pi_\Gamma(\bm{n}')/\partial \lambda = 0$, we have
\begin{equation}\label{eq:trade_off_deficiency_zero_CRNs1}
    \frac{\partial \langle n_i \rangle_\Gamma}{\partial \lambda}
    = \sum_{\rho\in\mathcal{R}'} \langle n_i \mu_\rho(\bm{n},\bm{n}') \rangle_\Gamma \frac{\partial}{\partial \lambda} 
    \ln \frac{T_{{\bm \nu}_\rho^-}}{T_{{\bm \nu}_\rho^+}}
\end{equation}
with
\begin{equation}
    \langle n_i \mu_\rho(\bm{n},\bm{n}') \rangle_\Gamma
    = \sum_{\bm{n},\bm{n}'\in\Gamma}
    n_i \mu_\rho(\bm{n},\bm{n}')
    \pi_\Gamma(\bm{n}) \pi_\Gamma(\bm{n}').
\end{equation}
The correlation function $\langle n_i \mu_\rho(\bm{n},\bm{n}') \rangle_\Gamma$ measures the fluctuations in the numbers of chemical species through the definition of $\mu_\rho(\bm{n},\bm{n}')$.
Since all the columns are linearly independent, the matrix $\mathsf{S}'$ has a left inverse matrix $(\mathsf{S}')^{-1}$ satisfying $(\mathsf{S}')^{-1}\mathsf{S}' = \mathsf{I}$.
The correlation function $\langle n_i \mu_\rho(\bm{n},\bm{n}') \rangle_\Gamma$ can be written as
\begin{equation}\label{eq:left_inverse}
    \langle n_i \mu_\rho(\bm{n},\bm{n}') \rangle_\Gamma
    = \sum_{j \in \mathcal{S}} (\mathsf{S}')^{-1}_{\rho j} {\rm Cov}_\Gamma \{ n_i, n_j \}.
\end{equation}
Note that the fluctuation-response relation in \eqref{eq:FRR} can be restored by noticing
\begin{equation}
   \sum_{\rho\in\mathcal{R}'} (\mathsf{S})_{\rho j}^{-1} \ln \frac{[\bm{X}]^{\bm{\nu}_
   \rho^-}_{\rm ss}}{[\bm{X}]^{\bm{\nu}_
   \rho^+}_{\rm ss}} = \ln [X_j]_{\rm ss}.
\end{equation}

Applying the inequalities \eqref{eq:MTT_inequality1} and \eqref{eq:MTT_inequality2} to \eqref{eq:trade_off_deficiency_zero_CRNs1} together with \eqref{eq:left_inverse}, we obtain trade-offs between the response and fluctuations,
\begin{equation}
    \left| k_{\pm\rho} \frac{\partial \langle n_i \rangle_\Gamma}{\partial k_{\pm\rho}} \right|
    \leq \sum_{\sigma\in\mathcal{R}'} \left| \sum_{j\in\mathcal{S}}  (\mathsf{S}')^{-1}_{\sigma j} {\rm Cov}_\Gamma \{ n_i, n_j \} \right|
\end{equation}
and
\begin{equation}
    \left| \frac{\partial \langle n_i \rangle_\Gamma}{\partial B_\rho} \right|
    \leq \sum_{\sigma\in\mathcal{R}'}  \left| \sum_{j\in\mathcal{S}} (\mathsf{S}')^{-1}_{\sigma j} {\rm Cov}_\Gamma \{ n_i, n_j \} \right| \tanh\left(\frac{\mathcal{F}}{4}\right).
\end{equation}

\bibliography{paper.bib}

\end{document}